\documentclass[12pt,letterpaper]{article}
\usepackage{jheppub}
\usepackage{graphicx}
\input{epsf}
\usepackage{epsfig}
\usepackage{epstopdf}
\usepackage{calligra}
\usepackage{subfigure}
\usepackage{color}
\usepackage[normalem]{ulem}

\definecolor{purple}{rgb}{0.6, 0, 1}

\DeclareMathAlphabet{\mathcalligra}{T1}{calligra}{m}{n}
\DeclareFontShape{T1}{calligra}{m}{n}{<->s*[2.2]callig15}{}

\newcommand{\ldS}{\mathcal{L}}
\newcommand{\lAdS}{\ell}
\newcommand{\be}{\begin{equation}}
\newcommand{\ee}{\end{equation}}
\newcommand{\bea}{\begin{eqnarray}}
\newcommand{\eea}{\end{eqnarray}}
\newcommand{\beas}{\begin{eqnarray*}}
\newcommand{\eeas}{\end{eqnarray*}}

\newcommand{\equ}[1]{\begin{equation} #1 \end{equation}}
\newcommand{\ali}[1]{\begin{align} #1 \end{align}}
\newcommand{\p}{\partial}

\DeclareMathOperator\csch{csch}
\DeclareMathOperator\tr{tr}
\DeclareMathOperator\sech{sech}
\DeclareMathOperator\arccot{arccot}

\title{Equivalence of Emergent de Sitter Spaces from Conformal Field Theory}
\author[1]{Curtis T.~Asplund,}

\author[1,2]{Nele Callebaut,}

\author[1]{and Claire Zukowski}

\affiliation[1]{Department of Physics, Columbia University, 538 West 120th Street, New York, NY 10027} 

\affiliation[2]{Instituut voor Theoretische Fysica, K.U. Leuven,
	Celestijnenlaan 200D B-3001 Leuven, Belgium} 

\emailAdd{ca2621@columbia.edu} 
\emailAdd{nmc2159@columbia.edu} 
\emailAdd{cez2103@columbia.edu}

\abstract{Recently, two groups have made distinct proposals for a de Sitter space that is emergent from conformal field theory (CFT). The first proposal is that, for two-dimensional holographic CFTs, the kinematic space of geodesics on a space-like slice of the asymptotically anti-de Sitter bulk is two-dimensional de Sitter space (dS$_2$), with a metric that can be derived from the entanglement entropy of intervals in the CFT. In the second proposal, de Sitter dynamics emerges naturally from the first law of entanglement entropy for perturbations around the vacuum state of CFTs. We provide support for the equivalence of these two emergent spacetimes in the vacuum case and beyond. In particular, we study the kinematic spaces of nontrivial solutions of $3$d gravity, including the BTZ black string, BTZ black hole, and conical singularities. We argue that the resulting spaces are generically globally hyperbolic spacetimes that support dynamics given boundary conditions at future infinity. For the BTZ black string, corresponding to a thermal state of the CFT, we show that both prescriptions lead to an emergent hyperbolic patch of dS$_2$. We offer a general method for relating kinematic space and the auxiliary de Sitter space that is valid in the vacuum and thermal cases. 
}

\begin{document}
\maketitle

\section{Introduction}

The AdS/CFT correspondence provides a powerful equivalence between a theory of (quantum) gravity in asymptotically anti-de Sitter space (AdS) and a conformal field theory (CFT) in one lower dimension. In spite of great progress in our understanding of the duality over the last two decades, the fundamental question of how bulk geometry emerges from the field theory has not been fully answered. Within the nascent  ``emergent spacetime from entanglement" program, a key tool for addressing this has been the Ryu-Takayanagi formula, which relates entanglement entropy in the CFT to the a\-re\-as of boundary-anchored bulk extremal surfaces \cite{Ryu:2006bv, Ryu:2006ef, Hubeny:2007xt}. The growing consensus is that at least outside ``shadow" regions blocked from Ryu-Takayanagi surfaces by barriers \cite{Freivogel:2014lja, Engelhardt:2015dta, Lin:2015lfa}, boundary entanglement entropy reconstructs the bulk spacetime and Einstein's equations \cite{Maldacena:2001kr, Ryu:2006bv, VanRaamsdonk:2010pw, Lashkari:2013koa, Faulkner:2013ica, Freedman:2016zud}.

One unforeseen consequence of this program was the identification of an auxiliary Lorentzian geometry from CFT entanglement data, distinct from the usual bulk AdS space. In fact, there have recently been two distinct proposals for an emergent de Sitter space from CFT. 

The first approach is rooted in an attempt to obtain the discretized geometry of a space-like slice of AdS from a boundary tensor network (ansatz for the ground state wavefunction) known as the Multi-Scale Entanglement Renormalization Ansatz (MERA) \cite{Swingle:2009bg, Nozaki:2012zj}. One school of thought has pointed out several challenges to the consistency of AdS/MERA, e.g., \cite{Bao:2015uaa}, and the authors of \cite{Czech:2015qta, Czech:2015kbp} proposed that the MERA tensor network is actually a discretization of a ``kinematic space" of boundary-anchored geodesics contained within a space-like slice of AdS$_3$, rather than of the bulk slice itself. Importantly, unlike the two-dimensional hyperbolic geometry ($H_2$) of a constant time slice of AdS$_3$, the geometry of this emergent kinematic space is Lorentzian. Specifically, it is a two-dimensional de Sitter space dS$_2$. The prescription in \cite{Czech:2015qta, Czech:2015kbp}, which matches earlier results related to differential entropy \cite{Balasubramanian:2013lsa, Myers:2014jia, Headrick:2014eia}, calculates the kinematic space metric solely from boundary entanglement entropy of intervals in the CFT (eq.~\eqref{kinematicprop}).

In a second construction that we refer to as the ``auxiliary dS prescription'' \cite{deBoer:2015kda}, a de Sitter space propagator is recognized hidden within the expression for the modular Hamiltonian of a $d$-dimensional CFT in the vacuum with a ball-shaped entangling region (eq.~\eqref{Hmod}). By applying the entanglement first law for small perturbations around the vacuum, the authors demonstrate that the entanglement perturbations satisfy a Klein-Gordon equation in an auxiliary $d$-dimensional de Sitter space. Unlike the kinematic space proposal, this construction is intrinsically dynamical. It applies in arbitrary dimensions, and as the authors stress, it is independent of the standard AdS/CFT correspondence.

As a maximally symmetric solution to Einstein's equation in the vacuum, it is perhaps not surprising to see de Sitter appear in different arenas. A priori, the two constructions need not be related beyond the vacuum case. It is thus a nontrivial check to see if the kinematic space and auxiliary dS prescriptions agree for bulk $3$d gravity solutions outside of pure AdS.\\

\noindent {\bf Our Results}: We apply the kinematic space prescription to additional nontrivial solutions of $3$d gravity: the (1-sided) BTZ black string, its quotient, and the conical singularity geometry. In the case of the quotiented BTZ black hole, we find good agreement with the partial results of \cite{Czech:2015xna, Czech:2015kbp} for the two-sided case (see Appendix~\ref{BTZAppendix}) and extend the analysis to include the effect of phase transitions in the entanglement entropy, which we show result in defects in kinematic space. In the other cases, the results are entirely new. For the BTZ black string and the conical singularities, we show that the resulting kinematic spaces are, respectively, the hyperbolic patch and glued together ``sub-de Sitter spaces" of de Sitter space, which are depicted in figures~\ref{BTZStringSubregion} and~\ref{CSSubregion}. The geodesics in these spacetimes can be mapped by large diffeomorphisms to parent geodesics in AdS, and for this reason their kinematic spaces are all subregions of the original de Sitter space. What is less obvious is that these kinematic spaces are not just subregions, but are \emph{causally well-behaved} spacetimes in their own right. Indeed, we present arguments that they are globally hyperbolic spacetimes whose boundary conditions can be set at future infinity. 

The global hyperbolicity of these kinematic subregions suggests that the space of geodesics can be interpreted as a background spacetime on which dynamical fields can propagate. It provides an immediate consistency check for the matching with the auxiliary de Sitter proposal, which is defined intrinsically in terms of dynamics and propagation. Furthermore, the fact that the kinematic spaces always have a Cauchy surface close to future infinity is consistent with the boundary conditions required for a boundary-to-bulk propagator.

For the BTZ black string example, we construct an explicit nontrivial match between the two emergent spacetimes. We demonstrate that in a direct extension of the result of \cite{deBoer:2015kda}, the modular Hamiltonian associated with a thermal CFT$_2$ interval is equal to the integral of the energy density times a Klein-Gordon propagator on the hyperbolic patch of dS$_2$ (eq.~\eqref{thermalHprop}). This patch is precisely the kinematic space we obtain in Section~\ref{BTZSec}. The matching informs our formulation of the equivalence in Section~\ref{prescriptions}. It also suggests a refinement of the kinematic space prescription for a CFT on a cylinder from the entanglement entropy $S_{\rm ent}(u,v)$ of a CFT interval	$[u,v]$, with a length scale $\ldS$ explicitly reinstated on the right hand side: 
\ali{
	ds^2_{\text{$\mathcal K$ of CFT on cylinder}} &= \frac{12}{c} \ldS^2 \frac{\p^2 S_{\rm ent}(u,v)}{\p u \p v} du dv \qquad \text{with $\ldS = \mathcal S/\pi$}.  \label{refinedprescription}
}
The length scale corresponds to the de Sitter radius $\ldS$, which we fix to be the circumference $\mathcal S$ of the cylinder's compact dimension over $\pi$. \\

\noindent {\bf Outline}: The paper is organized as follows. In Section~\ref{KinSpace}, we first review the kinematic space prescription in the case of pure AdS. We use the prescription to derive the kinematic space of the BTZ black string, black hole and the conical singularity, and discuss the causal structure of the resulting spacetimes (Section~\ref{Causal}). (For a review of the $3$d gravity solutions and the embedding coordinates and Penrose transformations used, see Appendices~\ref{Background} and~\ref{EmbeddingPenrose}, and for the relation to existing work on the BTZ quotient see Appendix~\ref{BTZAppendix}.) In Section~\ref{AuxSpace}, we review the auxiliary de Sitter construction and extend the construction to the thermal state, providing a match with the results of Section~\ref{BTZSec}. In Section~\ref{discussion}, we summarize our proposed refinement of the two constructions. We conclude with a discussion of our results and areas for future work.\\

\noindent In the final stages of preparing this paper, we learned of other upcoming results~\cite{Bartektalk, deBoertalk} that overlap with our work.

\section{Kinematic Space}\label{KinSpace}

Kinematic space, as formulated in \cite{Czech:2015qta, Czech:2015kbp}, can be defined for any CFT in any state. However, our main interest is in two-dimensional holographic CFTs which have an asymptotically AdS$_3$ bulk spacetime, where the kinematic space has a geometric interpretation as a space of boundary-anchored, oriented geodesics. 

Given any time-reflection symmetric asymptotically AdS$_3$ spacetime, there is a time coordinate $t$ such that all space-like extremal curves that anchor on boundary points with $t=0$ are entirely confined to a space-like slice defined by the condition $t=0$. This follows directly from the reflection symmetry about $t=0$ and includes, in particular, static spacetimes, which would have this property at each $t$. 
A constant time slice of a locally AdS$_3$ bulk will have a 2-dimensional hyperbolic geometry $H_2$, depicted in figures \ref{PoincareDiskAdS} and \ref{PoincareDiskBTZ}. 

The Ryu-Takayanagi holographic entanglement entropy proposal states that the entanglement entropy of an interval $[u,v]$ in the CFT (at $t=0$) is proportional to the length of the (unique) boundary-anchored geodesic with minimal length that is homologous to the interval \cite{Ryu:2006bv, Hubeny:2007xt, Headrick:2007km}. We take kinematic space to refer to the set of these geodesics for all intervals of the CFT.

\begin{figure}[t!]
	\centering
	\includegraphics[width=4.5in]{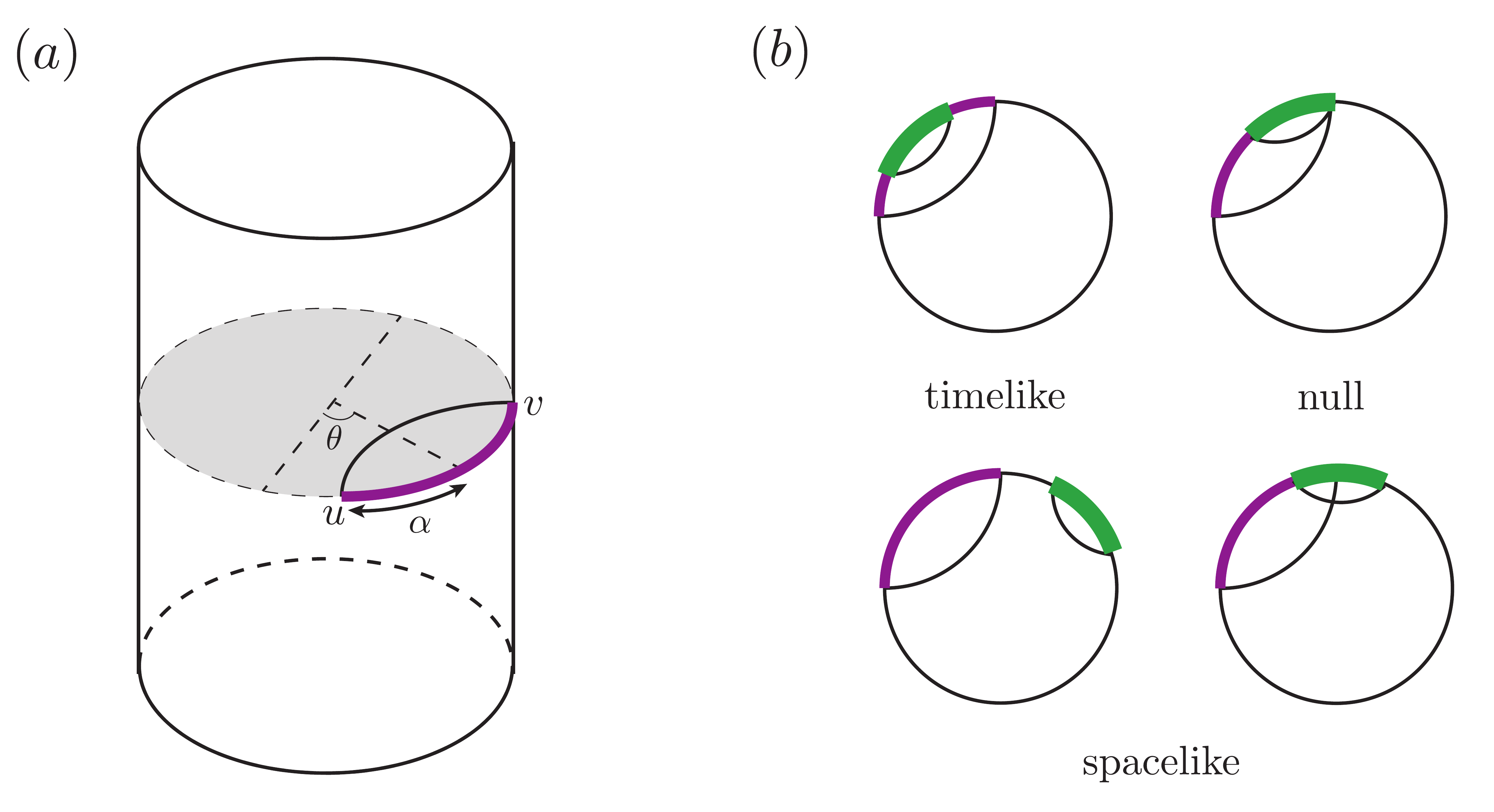}
	\caption{(a) A boundary-anchored geodesic on a constant time slice of AdS, with boundary interval in thick purple. The interval or equivalently the geodesic anchored to its endpoints is parametrized by the coordinates of its endpoints, $u$ and $v$, or by its midpoint angle $\theta$ and opening angle $\alpha$. (b) These geodesics enjoy a natural causal structure based on the containment relations of their boundary intervals (colored in thick purple and thicker green): geodesics are time-like separated if they have embedded boundary intervals (top left), null separated if they share a left or right endpoint (top right), and space-like separated if their boundary intervals are not embedded (bottom two). 
		}
	\label{KinematicSpace}
\end{figure}

By invoking results in integral geometry, \cite{Czech:2015qta} proposes a \emph{kinematic space prescription} for deriving a metric on kinematic space entirely from the entanglement entropy $S_{\rm ent}(u,v)$ of the boundary intervals:\footnote{We will in practice use the form given in \eqref{refinedprescription}. }
\be ds^2 = \frac{\partial^2 S_{\rm ent}(u,v)}{\partial u \partial v} du dv \label{kinematicprop}~.\ee 
This spacetime is Lorentzian due to a natural causal structure inherited from the containment relation of boundary intervals: two geodesics contained within one another are time-like separated, otherwise they are space-like separated or, in the marginal case where they share a left or right endpoint, null (see figure~\ref{KinematicSpace}). The null coordinates in kinematic space are the boundary endpoint coordinates $u$ and $v$. The geodesics or equivalently their boundary intervals can also be specified by the coordinates $\theta$ and $\alpha$ shown in figure~\ref{KinematicSpace}, representing the midpoint angle $\theta$ and the opening angle $\alpha$ of the geodesic. These are related to the endpoint coordinates by
\bea
u &=& \theta-\alpha~, \label{ucoord}  \\
v &=& \theta+\alpha~. \label{vcoord}
\eea 

\noindent {\bf A comment about notation}: \label{notationcomment} We will use the $(u,v)$ and $(\theta,\alpha)$ coordinates of figure~\ref{KinematicSpace} to label a CFT interval, both when it is defined on a compact direction and when it is not. 
In the case of a compact, circular direction, we will use the convention that $[u,v]$, with $u \leq v$ and $u, v \in [0,2\pi]$, signifies the interval going counterclockwise from the point with angular coordinate $u$ to the point with angular coordinate $v$. By $[v,u]$, with $v > u$, we mean the closure of the complement of $[u,v]$. Note that we stick to a single orientation for intervals on the circle, while we will consider geodesics with both clockwise and counterclockwise orientations.
We will have non-compact directions in the discussions of the Poincar\'e patch of AdS and the BTZ black string, in which cases the corresponding $u,v,\theta,\alpha$ coordinates are not angular but have dimensions of length. Throughout, the coordinates $u$ and $v$ will be lightlike kinematic space coordinates, while $\alpha$ and $\theta$ will define time-like and space-like coordinates, respectively. Their detailed specification depends on the CFT state under consideration.

\begin{figure}[t!]
	\centering
	\includegraphics[width=6in]{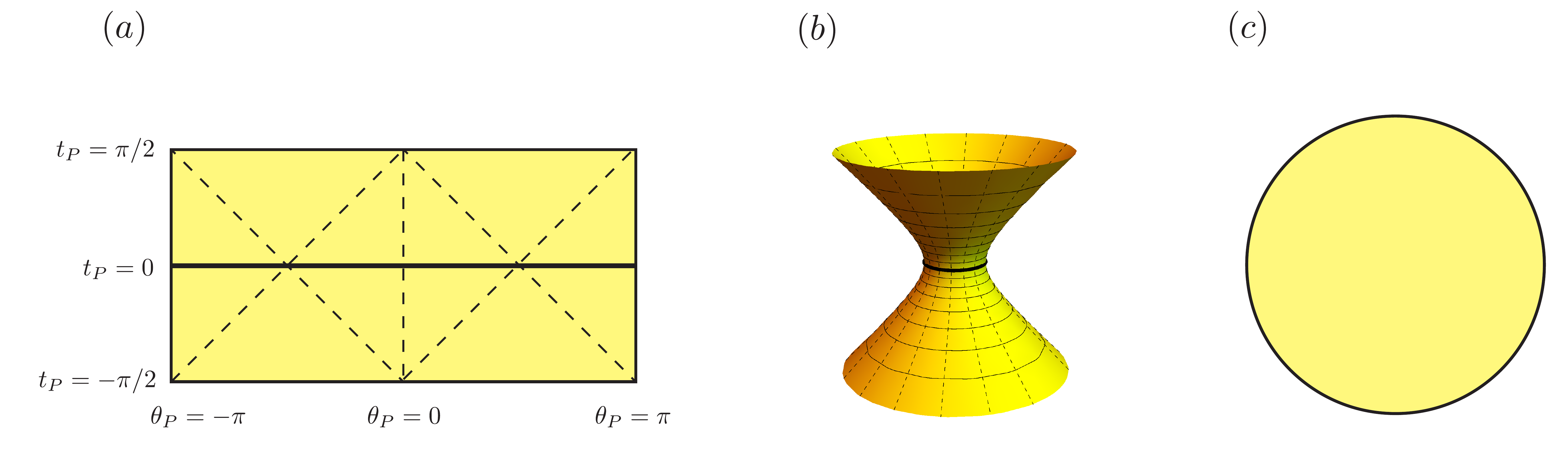} 
	\caption{ 
		The kinematic space for pure AdS$_3$ is a 2-dimensional de Sitter space, represented (a) as a Penrose diagram, using coordinates defined in eqs. \eqref{B8} and \eqref{B9}, and (b) as a dS$_2$ hyperboloid embedded in flat space $R^{1,2}$, with constant $t$ lines (black) and constant $\theta$ lines (dashed). The dS$_2$ waist at $\alpha = \pi/2$ is highlighted in thick black to stress that kinematic space is the space of \emph{oriented} geodesics: the entire expanding portion of de Sitter (above the waist) maps to all $H_2$ geodesics with one orientation, while the contracting region (under the waist) maps to the same geodesics but with opposite orientation. The geodesics cover the full constant time slice of AdS$_3$, represented in (c) as a Poincar\'e disk (cf.~figure \ref{PoincareDiskAdS}). 
			}
	\label{AdSSubregion}
\end{figure}

\subsection{Global AdS$_3$}\label{globalads}
We begin by reviewing the results of the prescription in the case of pure AdS$_3$. The dual is the vacuum state of a CFT on a cylinder with compact space-like dimension of circumference $\Sigma = 2\pi \mathcal R$ (see Appendix~\ref{Background} for a review of the conformal boundary). 
The boundary entanglement entropy of an interval of length $L = \mathcal R (v-u)$ is \cite{Calabrese:2004eu}
\be \label{eq:EEcirc} S_{\rm ent} = \frac{c}{3}\log{\frac{\Sigma}{\epsilon \pi} \sin{\frac{\pi L}{\Sigma}}}~,\ee
where $\epsilon$ is the UV cutoff, and $c$ is the central charge of the CFT, related to the AdS radius $\lAdS$ and 3-dimensional gravitational constant $G_3$ through~\cite{Brown:1986nw}
\ali{
	c = \frac{3 \lAdS}{2 G_3}~. \label{cofG} 
	}
We note that the formula \eqref{eq:EEcirc} is universal, i.e., it only depends on the particular CFT through the central charge.

\begin{figure}[t!]
	\centering 
	\includegraphics[width=4.5in]{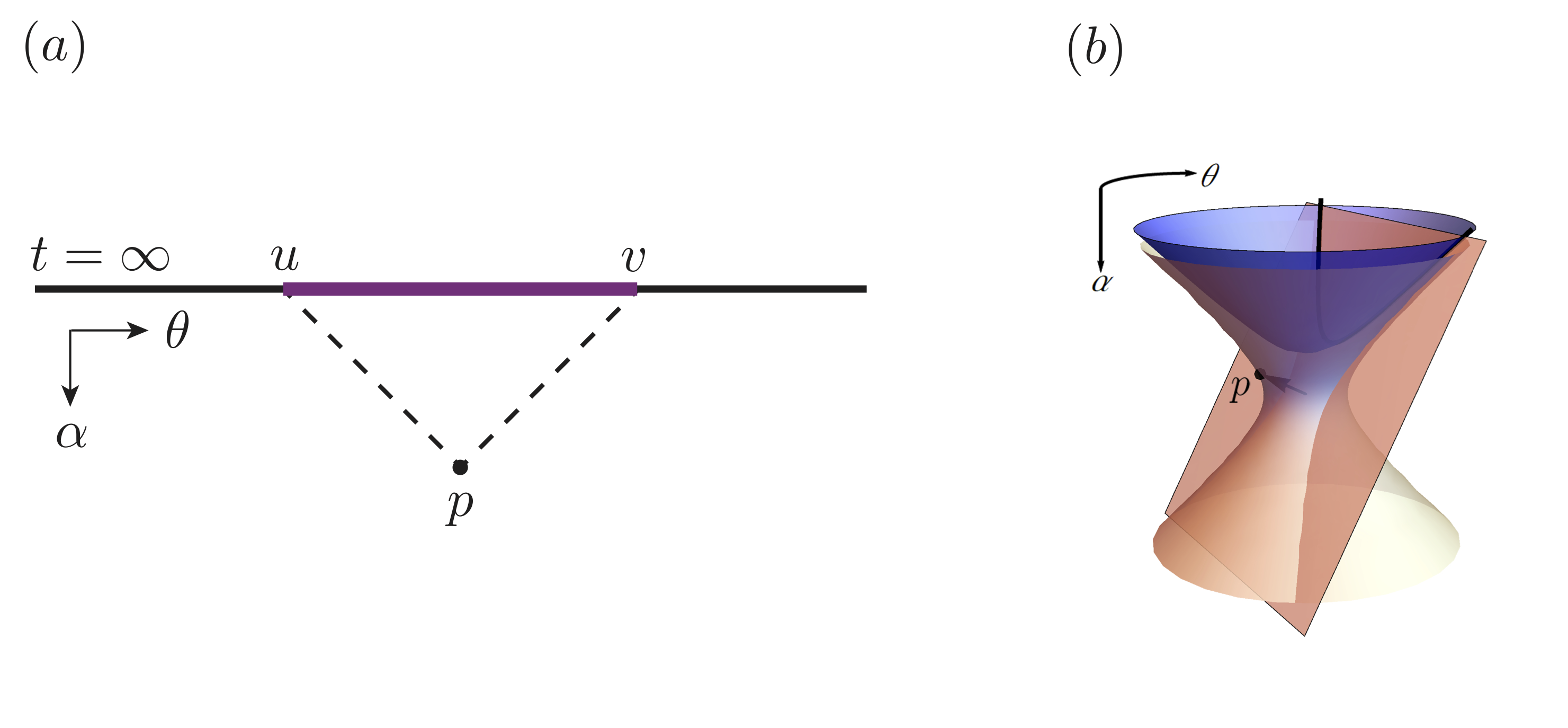}
	\caption{Kinematic space defines an emergent dS$_2$ on a CFT$_2$ by associating a point $p$ on de Sitter to a given CFT interval $[u,v]$, or equivalently, to the corresponding boundary-anchored $H_2$ geodesic $[u,v]$. (a) The CFT interval lies at the asymptotic future boundary of dS$_2$, and can be identified with a point $p$ in de Sitter at the tip of the lightcone extending into the bulk. (b) In embedding space, geodesics on $H_2$ are intersections of origin-centered planes with the $H_2$ hyperboloid (blue). Each such plane specifies a point in the dS hyperboloid via its outward pointing normal.}
	\label{Embedding}
\end{figure}

From the prescription \eqref{refinedprescription}, we find 
\be ds^2 = \ldS^2 \frac{du dv}{\sin^2{\left(\frac{v-u}{2}\right)}} \qquad (0 \leq u,v\leq 2\pi)~.\ee
This is the metric of a $2$-dimensional de Sitter space dS$_2$ with radius $\ldS$, in conformally compactified null coordinates.
In the coordinates defined in \eqref{ucoord}-\eqref{vcoord}, it takes the form 
\be ds^2 = \frac{\ldS^2}{\sin^2\alpha} (-d\alpha^2 + d\theta^2) \qquad ( 0 \leq \alpha \leq \pi, \ 0 \leq \theta \leq 2\pi)~.\label{globalcoords}\ee
In other words, the opening angle $\alpha$ of the geodesic is a natural time coordinate on the space of geodesics.

To observe which portion of de Sitter is covered by these coordinates, we convert to global coordinates by the transformation
\be t = - \log{\tan{\frac{\alpha}{2}}}~. \label{tOfalphaglobal} \ee
The metric becomes
\be ds^2 = \mathcal{L}^2\left(-dt^2 +\cosh^2{t} \, d\theta^2) \qquad (-\infty < t < \infty, \ 0 \leq \theta \leq 2\pi \right)~. \label{AdSmetric}\ee 
By the embedding \eqref{globalembedding}, 
this covers the full de Sitter hyperboloid \eqref{dShyperboloid} shown in figure~\ref{AdSSubregion}b. The expanding and contracting portions are equivalent up to orientation: a geodesic $(\theta,\alpha)$ in the region above the waist $\alpha = \frac{\pi}{2}$ corresponds to a geodesic below the waist with the same bulk profile but opposite orientation, $(\theta + \pi, \pi - \alpha)$. The two orientations correspond to complementary boundary intervals $[u,v]$ and $[v,u]$, which in a pure state share the same Ryu-Takayanagi curve and boundary entanglement entropy.

As presented in figure~\ref{Embedding}a, a boundary interval $[u,v]$ is mapped to a point $p$ in dS$_2$ at the tip of the lightcone that projects to the interval.  
Geodesics with zero opening angle correspond to points on the boundary, hence the conformal boundary can be identified with the asymptotic future $\mathcal I_+$ ($t \rightarrow \infty$) of kinematic space, which is the topmost line in the Penrose diagram shown in figure~\ref{AdSSubregion}a.
The mapping between geodesics on $H_2$ and points in dS$_2$ is also intuitive from the embedding diagram, since geodesics on $H_2$ are located at its intersection with a plane centered at the origin. 
These map to two points in de Sitter space via the two normal vectors of the plane: one on the expanding region (see figure~\ref{Embedding}) and one on the contracting region of de Sitter. Such geodesics share the same radial profile but have opposite orientation.

Because of the universality of the entropy formula in eq.~\eqref{eq:EEcirc}, the kinematic space we have rederived here is also universal for all CFTs, not just holographic ones where $c$ is related to parameters of a holographically dual bulk geometry by eq.~\eqref{cofG}.

\subsection{Poincar\'e Patch}\label{PoincareSec}

\begin{figure}[t!]
	\centering
	\includegraphics[width=6in]{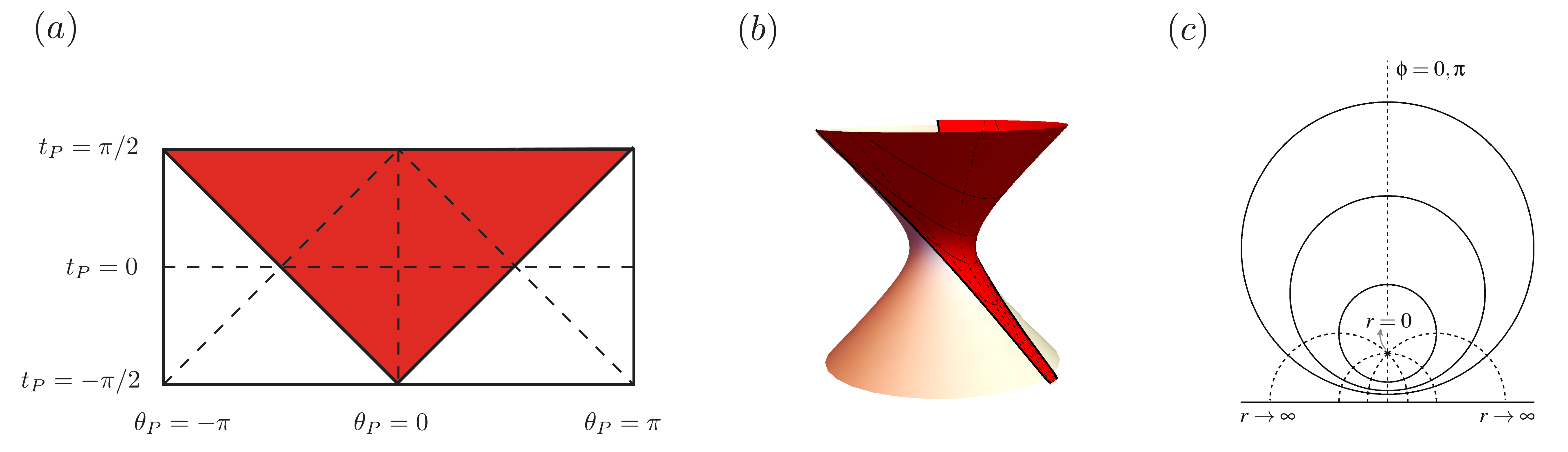} 
	\caption{The kinematic space for the Poincar\'e patch of AdS$_3$ is the planar patch of dS$_2$, depicted in red in (a) the Penrose diagram, using coordinates defined in eqs.~\eqref{PlanarPenrose1} and \eqref{PlanarPenrose2}, and (b) the embedding diagram, with lines of constant time (solid) and constant $\theta$ coordinates (dashed). The geodesics cover the full constant time slice of AdS$_3$, represented in (c) as the half-plane (cf.~figure \ref{PoincareDiskAdS}).}
	\label{PoincareSubregion}
\end{figure}

If we consider the Poincar\'e patch of the bulk AdS, its conformal boundary will be a plane instead of a cylinder. In that limit, $\Sigma \gg L$, the single-interval boundary entanglement entropy \eqref{eq:EEcirc} will reduce to
\ali{
	S_{\rm ent} = \frac{c}{3} \log \frac{L}{\epsilon} = \frac{c}{3} \log \frac{v-u}{\epsilon}~,
	}
where $u$ and $v$ are again the interval endpoints, but now with a dimension of \emph{length}. The corresponding kinematic space metric is
\ali{
	ds^2 = 4 \ldS^2 \frac{du \, dv}{(v-u)^2} \qquad (-\infty < u,v < \infty)~
	}
or, in the coordinates of the midpoint $\theta$ and ``radius'' $\alpha$ of the interval,
\ali{
	ds^2 = \frac{\ldS^2}{\alpha^2} (-d\alpha^2 + d\theta^2) \qquad (0 \leq \alpha < \infty,\ -\infty <  \theta < \infty)~. \label{Kplanar}
	}
This is the metric of a 2-dimensional de Sitter space with radius $\ldS$ in planar coordinates \eqref{planarembedding}, which cover the planar patch, see figure~\ref{PoincareSubregion}.\footnote{In all the cases we consider, we could equally well cover the patch connected to $\mathcal I^-$ rather than $\mathcal I^+$ by flipping the sign of the time coordinate. This would correspond to giving the geodesics an opposite orientation. }

\subsection{BTZ Black String}\label{BTZSec}

The BTZ metric,
\be ds^2 = -\left(\frac{r^2-r_+^2}{\lAdS^2} \right) dt^2 + \left(\frac{r^2-r_+^2}{\lAdS^2} \right)^{-1} dr^2 + r^2 d\phi^2~,\ee
is a nontrivial solution of $3$d gravity, where $\lAdS$ is the AdS radius and $r_+$ is the horizon radius, related to its temperature via $r_+ = 2\pi \lAdS^2/\beta$ \cite{Banados:1992wn, Banados:1992gq}. We refer to the geometry as the ``BTZ black string" in the unwrapped (covering space) case, when the $\phi$ coordinate ranges over $-\infty< \phi < \infty$, and as the ``BTZ black hole" when considering the quotient space, which restricts $\phi$ to $-\pi \leq \phi \leq \pi$. We discuss the BTZ black hole in the next subsection.

The BTZ black string is dual to a CFT in a thermal state at temperature $\beta^{-1}$, and so effectively lives on a Euclidean-signature cylinder with an infinite space dimension and a compact imaginary \emph{time} dimension of length $\beta$. The formula for the entanglement entropy of an interval $[u,v]$ of length $L = v-u$ is \cite{Calabrese:2004eu} 
\be 
\label{eq:EEtherm} 
	S_{\rm ent} = \frac{c}{3} \log{\left(\frac{\beta}{\epsilon\pi}\sinh{\frac{\pi L}{\beta}}\right)}
		= \frac{c}{3} \log{\left(\frac{\beta}{\epsilon\pi}\sinh{\frac{\pi (v-u)}{\beta}}\right)} \ 
	\ee
and, similarly to eq.~\eqref{eq:EEcirc}, it is universal.

We again apply eq.~\eqref{refinedprescription}, but with $u$ and $v$ taken to be \emph{lengths} rather than angles, parameterizing the endpoints of the interval along the infinite spatial boundary. We find that the metric on kinematic space is
\be ds^2 = \frac{4\pi^2 \ldS^2}{\beta^2} \frac{du dv}{\sinh^2{\left(\frac{\pi(v-u)}{\beta}\right)}} \qquad (-\infty < u,v < \infty) ~,\ee
or equivalently
\be ds^2 = \frac{4\pi^2 \ldS^2}{\beta^2} \frac{-d\alpha^2 + d\theta^2}{\sinh^2{\left(\frac{2 \pi \alpha}{\beta}\right)}} \qquad (0 \leq \alpha < \infty, \ -\infty < \theta < \infty)~ \label{KBTZstring} \ee 
in the coordinates defined in \eqref{ucoord}-\eqref{vcoord}. 

\begin{figure}[t!]
	\centering
	\includegraphics[width=6in]{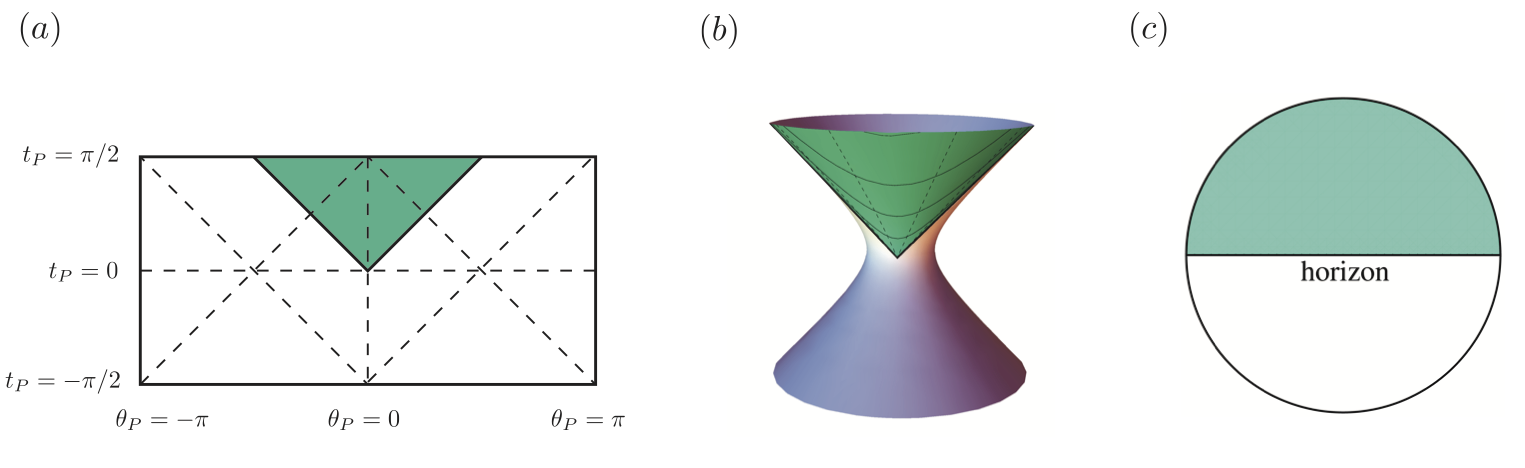} 
	\caption{The kinematic space for the 1-sided BTZ black string is the hyperbolic patch of de Sitter, depicted in green in (a) the Penrose diagram, using coordinates defined in eqs.~\eqref{HyperbolicPenrose1} and \eqref{HyperbolicPenrose2}, and (b) the embedding diagram with constant time lines (solid) and constant $\theta$ lines (dashed). (c) The geodesics cover one half of the Poincar\'e disk, that is, one outside-horizon region (cf.~figure \ref{PoincareDiskBTZ}).}
	\label{BTZStringSubregion}
\end{figure}

We can convert to hyperbolic coordinates via the transformation 
\be
\tau = - \log{\tanh{\frac{\pi \alpha}{\beta}}}~,\qquad \chi = \frac{2 \pi \theta}{\beta}~   \label{tOfalphahyp}.
\ee
Now, the metric takes the form
\be ds^2 = \ldS^2 (-d\tau^2 + \sinh^2{\tau} d\chi^2) \qquad (0 \leq \tau < \infty, -\infty < \chi < \infty)~.\label{BTZeqn}\ee
By the embedding \eqref{hyperbolicembedding}, these coordinates and ranges cover the hyperbolic patch of de Sitter, depicted in figure~\ref{BTZStringSubregion}. Note that this is only equal to  the hyperbolic patch of the AdS$_3$ kinematic space dS$_2$ when the de Sitter radii $\ldS$ are the same, which is not necessarily the case. Indeed, we will see in Section~\ref{AuxSpace} that $\ldS$ is temperature dependent in the BTZ case. 

The geodesics corresponding to the hyperbolic patch cover the region outside the horizon on the spatial BTZ slice and are homologous to CFT intervals on a single asymptotic boundary. Due to the boundary being in a mixed state, complementary intervals no longer correspond to geodesics with opposite orientation and indeed, such intervals are not included in kinematic space once we restrict to a single boundary.  Alternatively, the state can be represented as a pure state (the thermofield double state) on two copies of the CFT, which amounts to considering the 2-sided BTZ black string. \\

\subsection{BTZ Black Hole} \label{BTZqSec}

The BTZ black hole geometry can be obtained from the black string by quotienting by an appropriate subgroup of the AdS$_3$ isometry group \cite{Banados:1992gq, Brill:1998pr}. The boundary CFT effectively lives on a torus with spatial radius $\mathcal{R}$ and a compact imaginary time direction of length $\beta = 2\pi \ell \mathcal{R}/r_+$ (see Appendix~\ref{Background}). The quotient results in space-like geodesics with new global characteristics, including some that wind an arbitrary number of times around the circular horizon.

To parametrize these geodesics, we will again take the midpoint and opening angles $\theta, \alpha$ to be \emph{angular} coordinates along the compact spatial circle, related to the \emph{lengths} $\theta, \alpha$ of the BTZ string by a factor of the radius $\mathcal{R}$. The profile of a space-like BTZ geodesic is inherited from the associated minimal parent geodesic in the BTZ black string geometry, and is given by
\ali{
	r(\phi; \alpha, \theta)   &= r_+ \frac{\cosh \frac{2\pi \alpha \mathcal R}{\beta}}{\sqrt{\cosh^2 \frac{2\pi \alpha \mathcal R}{\beta} - \cosh^2(\frac{2\pi \phi \mathcal{R}}{\beta}-\frac{2\pi\theta \mathcal R}{\beta})}}  \qquad (\theta - \alpha \leq \phi  \leq \theta + \alpha)~. \label{rgeodesic}
}
For a given geodesic with opening angle $\alpha$, the maximal extent it can reach into the bulk is $r_{\rm crit}(\alpha) =  r_+  \coth \frac{2\pi \alpha\mathcal R}{\beta}$. 

 \begin{figure}[t!]
	\centering
	\includegraphics[width=4.5in]{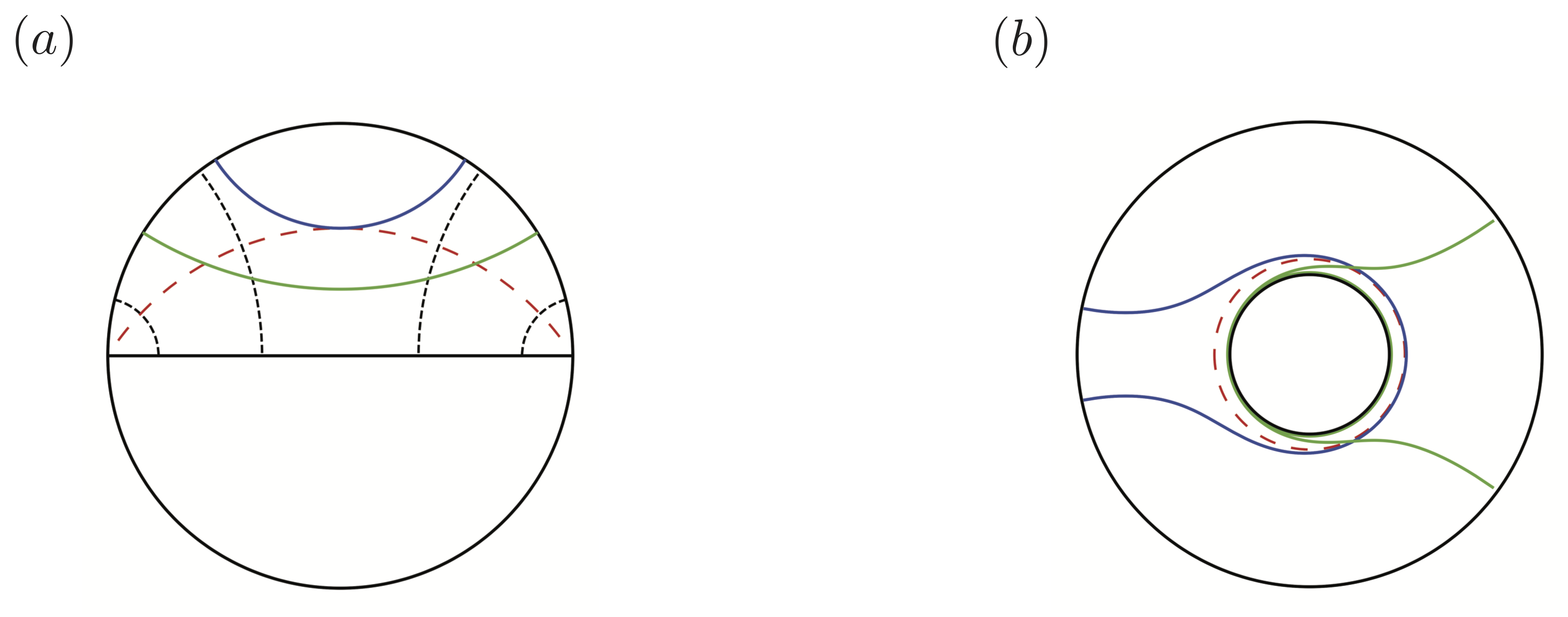} 
	\caption{
		(a) Two example geodesics on the covering BTZ black string slice and (b) their counterparts on the slice of the quotiented BTZ black hole geometry. The covering space consists of an infinite number of copies of the fundamental domain $-\pi \leq \phi < \pi$, marked by black dashed boundaries. The blue geodesic is non-winding ($\alpha < \pi$) with maximal opening angle $\alpha_c$, defined in eq.~\eqref{alphac}, and touches the entanglement shadow (in dashed red) defined in eq.~\eqref{entshadowbh}. The green geodesic with $\alpha > \pi$ winds around the singularity once and enters the entanglement shadow region. (The plots were made for $\beta/\mathcal{R} = 2\pi$ and $l=1$.) }
	\label{BTZgeodesics}
\end{figure} 

In terms of the spatial slice we are considering, the quotient amounts to identifying the constant $\phi$ lines bounding the fundamental domain $-\pi \leq \phi < \pi$ in the Poincar\'e disk representation of the slice (see Appendix~\ref{Background} and figure~\ref{PoincareDiskBTZ}). 
Depending on whether a BTZ geodesic $(\theta,\alpha)$ with $-\pi \leq \theta \leq \pi$ has a width less than the angular width $\pi$ of a fundamental domain, crosses over multiple identifications or coincides with the horizon, the respective images of the geodesics in the quotient space will be non-winding, winding or infinitely winding. The geodesic is non-winding when $\alpha \leq \pi$ and $n$ times winding when $n \pi \leq \alpha < (n+1)\pi$ (with $0 \leq n < \infty$). See figure~\ref{BTZgeodesics} for some example geodesics in the covering space and its quotient.

\begin{figure}[t!]
	\centering
	\includegraphics[width=2in]{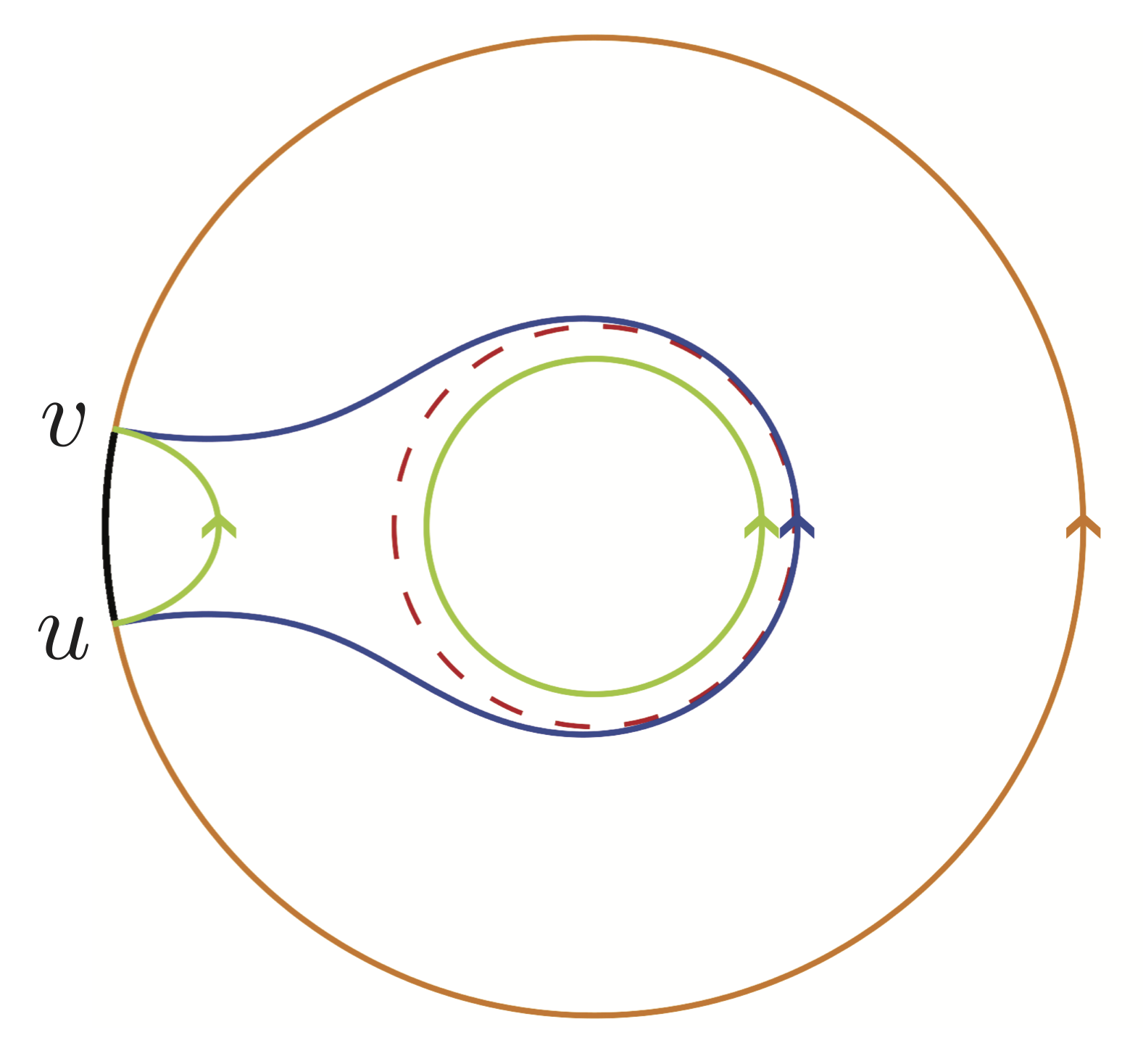} 
	\caption{
		Constant time slice with orange interval $[u,v]$. When the interval opening angle is larger than the critical angle $\alpha > \alpha_c$ (with $\pi/2 < \alpha_c < \pi$), the homologous, connected, non-winding geodesic in blue (with counterclockwise orientation) has larger length than the disconnected geodesic in green. The latter is homologous to the orange interval if the horizon-wrapping geodesic has counterclockwise orientation and the non-wrapping boundary-anchored geodesic runs from $u$ to $v$ (counterclockwise). The boundary-anchored portion of this disconnected geodesic is equal to the minimal boundary-anchored geodesic for the complementary interval $[v,u]$ with opening angle $\pi - \alpha$ but with the orientation reversed, so that together with the horizon-wrapping geodesic it is homologous to the big orange interval instead.
	} \label{BTZphasetransition} 
\end{figure}

The geodesics that compute entanglement entropy are the minimal, homologous ones, which we refer to as Ryu-Takayanagi geodesics. For sufficiently small intervals, the Ryu-Takayanagi geodesics are the non-winding geodesics whose length is given by eq.~\eqref{eq:EEtherm}. However, there exists a critical size past which there is a new family of disconnected geodesics that have smaller length than the connected homologous ones \cite{Headrick:2007km}. 
This can be interpreted as a ``phase transition" in the entanglement entropy---see \cite{Hubeny:2013gta, Chen:2014hta, Chen:2015kua} for recent studies. 
The disconnected geodesics consist of the disjoint union of a horizon-wrapping geodesic (obtained from eq.~\eqref{rgeodesic} in the $\alpha \rightarrow \infty$ limit) and the non-winding geodesic for the complementary interval, which has opening angle $\pi - \alpha$ and midpoint angle $\theta \pm \pi$ (i.e., interchanged endpoints $u$ and $v$). This is illustrated in figure \ref{BTZphasetransition}.

\begin{figure}[t!]
	\centering 
	\includegraphics[width=2.7in]{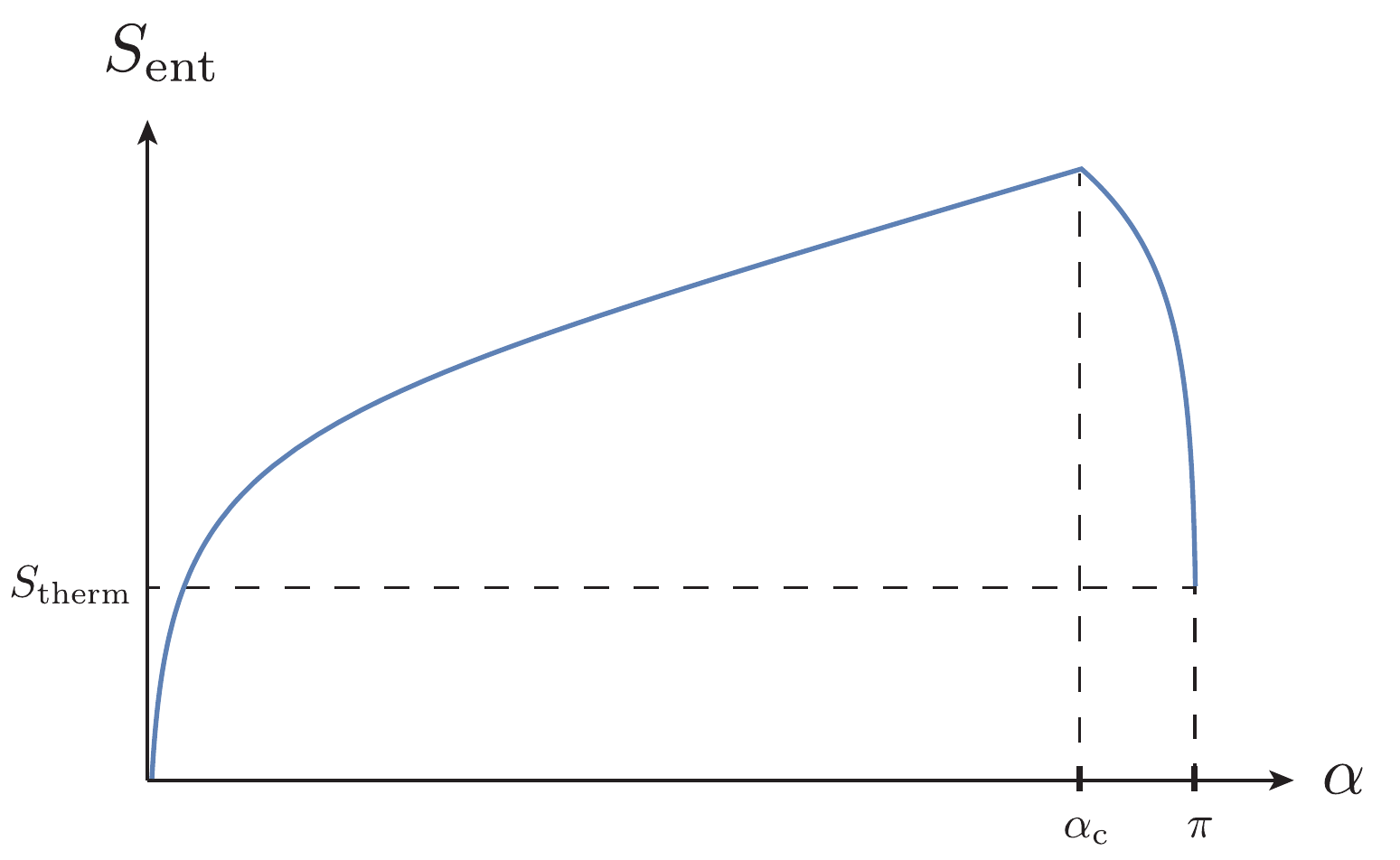} 
		\caption{The entanglement entropy of boundary intervals for the BTZ black hole, as a function of the opening angle $\alpha$ of the interval. At the critical angle $\alpha_{\rm c}$, there is a phase transition and a new family of disconnected Ryu-Takayanagi geodesics have minimal length. For very large interval size, the entanglement entropy approaches the thermal entropy.}
\label{Sent-bh} 
\end{figure}

The lengths of these geodesics compute the entanglement entropy of intervals, which is piecewise defined as
\ali{
S_{\rm ent} = \left\{ \begin{array}{ll} \frac{c}{3} \log \left( \frac{\beta}{\epsilon \pi} \sinh \frac{2\pi \alpha \mathcal{R}}{\beta} \right) \qquad \qquad \qquad \qquad \alpha < \alpha_c~,  \\
\frac{c}{3} \frac{2\pi^2 \mathcal R}{\beta} + \frac{c}{3}  \log \left( \frac{\beta}{\epsilon \pi} \sinh \frac{2\pi ( \pi-\alpha)\mathcal R}{\beta} \right) \qquad \ \ \alpha > \alpha_c~, \end{array} \right.  
}
with a discontinuity in the first derivative at $\alpha_c$, the critical angle for the phase transition (see figure \ref{Sent-bh}).\footnote{Such discontinuities arise from considering classical gravity in the bulk, and will be smoothed out when $1/c$ corrections are taken into account~\cite{Hubeny:2013gta}.}

The critical angle is obtained from equating the contributions from each branch: 
\be \alpha_{\rm c} = \frac{\beta}{4\pi \mathcal R}\log \left(\frac{1}{2} + \frac{1}{2} \exp\left(\frac{4\pi^2 \mathcal R}{\beta}\right) \right)~,\label{alphac}\ee
with $\pi/2 < \alpha_{\rm c} < \pi$, which approaches $\pi$ in the high-temperature limit $\mathcal{R}/\beta \gg 1/2\pi$. 

The geometry exhibits a temperature-dependent entanglement shadow, since apart from the horizon $r=r_+$ the Ryu-Takayanagi geodesics cannot probe below the radius
\be r_{\rm max} = r_{\rm crit}(\alpha_{\rm c}) =  r_+  \coth \left\{ \frac{1}{2} \log \left(\frac{1}{2} + \frac{1}{2} \exp\left(\frac{4\pi^2 \mathcal R}{\beta}\right) \right)\right\}~. \label{entshadowbh} \ee
The maximal entanglement shadow occurs for the lowest allowed temperature $\mathcal{R}/\beta = 1/2\pi$ (the Hawking-Page phase transition temperature), below which the bulk geometry is thermal AdS rather than a BTZ black hole \cite{Hawking:1982dh}. It is illustrated in figure \ref{BTZgeodesics}. The smallest allowed shadow region occurs in the high-temperature limit $\mathcal{R}/\beta \gg 1$, when the Ryu-Takayanagi geodesics can reach all the way to the horizon.

\begin{figure}[t!]
	\centering 
	\includegraphics[width=5in]{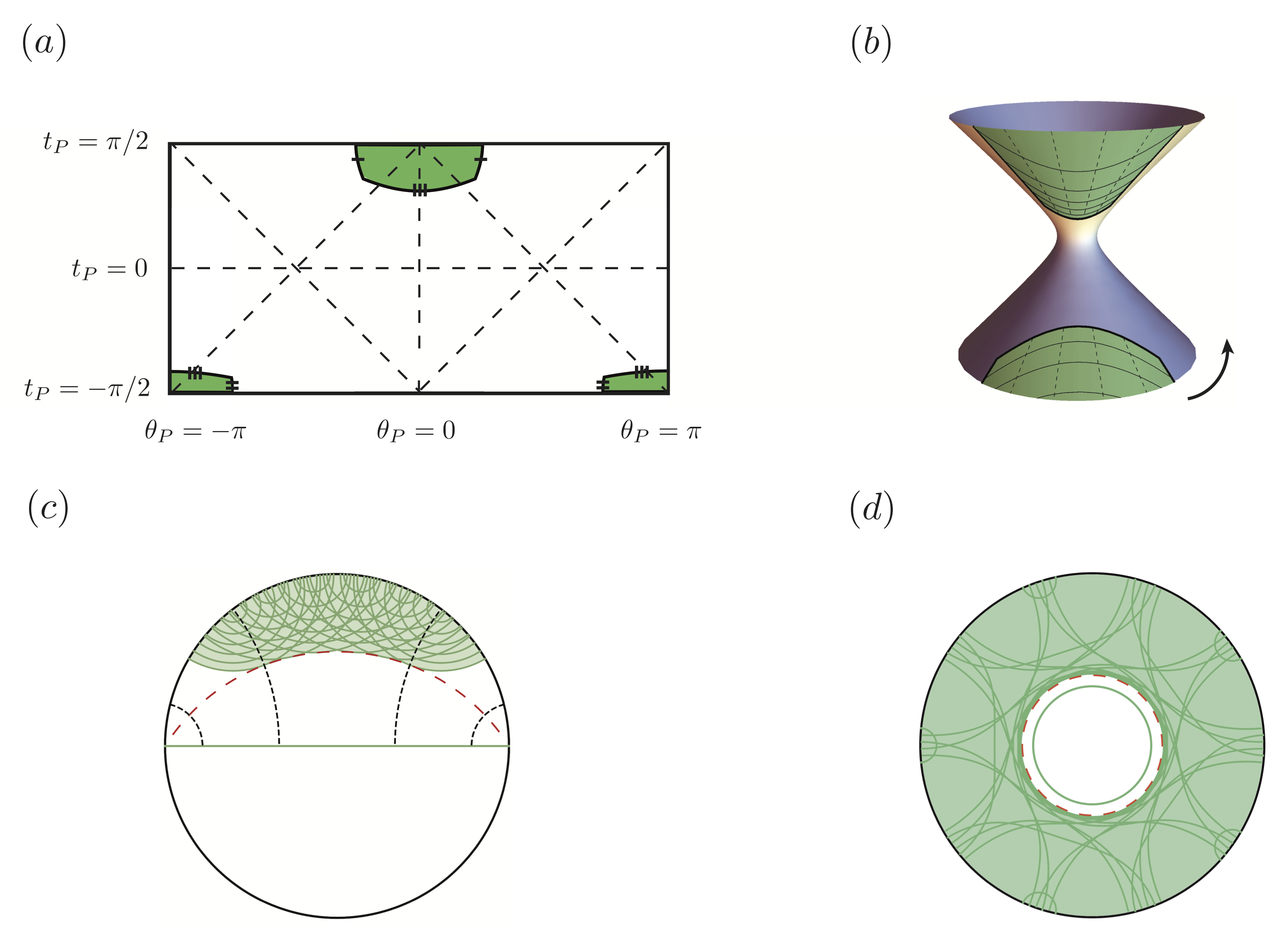} 
	\caption{
			The kinematic space for the BTZ black hole, shown in (a) the Penrose diagram, using coordinates defined in eqs.~\eqref{HyperbolicPenrose1} and \eqref{HyperbolicPenrose2}, and (b) the embedding diagram, with lines of constant time (solid) and constant $\theta$ coordinates (dashed). (The arrow in the embedding diagram indicates that the lower patch is actually located on the reverse side of the hyperboloid.) The space consists of two distinct subportions of the upper and lower hyperbolic patches of dS$_2$ corresponding to the two phases of Ryu-Takayanagi geodesics, which after a $\theta \rightarrow \theta+\pi$ rotation are glued together on their constant $\alpha=\alpha_{\rm c}$ boundaries, with a defect along the identification. The portion of the upper hyperbolic patch that is covered increases as a function of $\mathcal R/\beta$, and in the high-temperature limit $\mathcal R/\beta \gg 1/2\pi$ approaches 
			the full upper hyperbolic patch that is the kinematic space of the BTZ string, depicted in figure~\ref{BTZStringSubregion}. The Ryu-Takayanagi geodesics cover a region outside the entanglement shadow on (c) the covering BTZ black string slice and (d) the quotiented BTZ slice, depicted here for the lowest allowed temperature $\mathcal{R}/\beta = 1/2\pi$ (and $l=1$) which gives the maximal entanglement shadow region for the BTZ geometry. The minimal entanglement shadow region occurs in the high-temperature limit when the Ryu-Takayanagi geodesics can extend all the way up to the horizon $r = r_+$. 
			}
			\label{BTZqSubregion}
\end{figure}  

\begin{figure}[t!]
	\centering
	\includegraphics[width=5in]{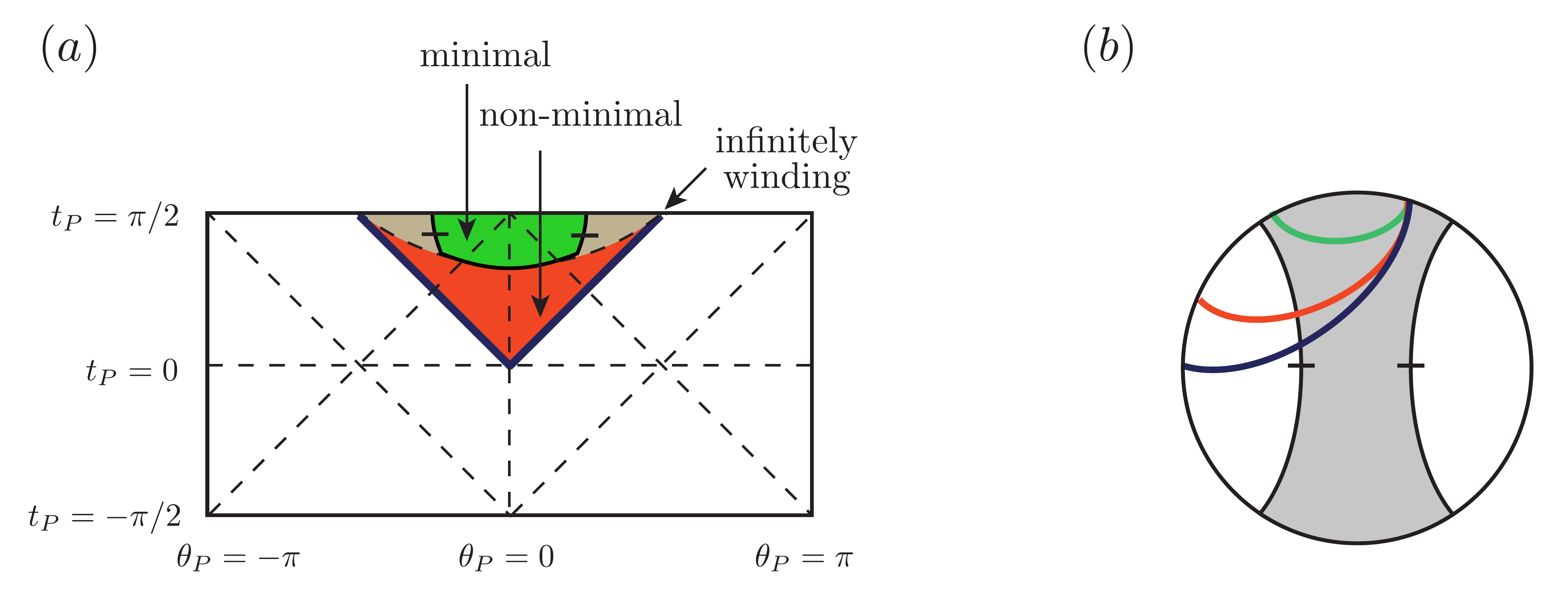} 
	\caption{
		(a) The Penrose diagram for the kinematic space of the covering BTZ black string, color-coded to show the regions of geodesics that upon quotienting to the black hole geometry either remain minimal (green and brown), become 
		non-minimal (orange), or become infinitely winding (dark blue).
		The innermost green region is the kinematic space of the BTZ black hole. Its constant $\theta$ boundaries are identified in the quotient, along with each subsequent boundary of (an infinite number of) fundamental domain copies $\theta \sim \theta + 2 \pi$ (in brown).  	The constant-$\alpha$ lower boundary at $\alpha = \alpha_{\rm c}$ marks the maximum opening angle before minimal geodesics become non-minimal. The non-minimal region contains both non-winding and (finitely) winding geodesics: a different constant-$\alpha$ boundary at even larger $\alpha= \pi > \alpha_{\rm c}$ (not drawn) would separate the non-minimal non-winding geodesics from the winding ones.
		(b) An example geodesic for each region depicted on the Poincar\'e disk. 
	} 
	\label{BTZBH1}
\end{figure}

As kinematic space is by definition constructed out of entanglement entropy data (see eq.~\eqref{kinematicprop}), it is not the space of \emph{all}  boundary-anchored, oriented geodesics in these quotiented geometries, but specifically the Ryu-Takayanagi geodesics.\footnote{Here we take a different point of view than in~\cite{Czech:2015xna, Czech:2015kbp}, where the non-minimal geodesics are included in `kinematic space'. To compare to their notion of kinematic space, it suffices to consider the kinematic space of the covering AdS, see figures \ref{BTZBH1} and \ref{con-colors}.} The kinematic space is given by:  

\be ds^2 = \frac{4\pi^2 \ldS^2 \mathcal{R}^2}{\beta^2} f(\alpha) (-d\alpha^2 + d\theta^2) \qquad \qquad (0 < \alpha < \pi, \, -\pi < \theta < \pi)~, \label{KBTZbh}\ee
where

\ali{
	f(\alpha) = \left\{ \begin{array}{ll}
		 \csch^2 \frac{2\pi \alpha \mathcal{R}}{\beta} &0 \leq \alpha < \alpha_c\\ 
		\frac{\beta}{\pi \mathcal{R}}\left(\tanh{\frac{\pi^2\mathcal{R}}{\beta}}+\coth{\frac{\pi^2\mathcal{R}}{\beta}}\right)\delta(\alpha-\alpha_{\rm c})  \qquad &\alpha = \alpha_c\\
		\csch^2 \frac{2\pi(\pi  - \alpha)\mathcal R}{\beta}   &\alpha_c < \alpha \leq \pi
	\end{array}  \right. \,.
	}
The delta function in the metric at $\alpha = \alpha_{\rm c}$ results from the discontinuity in the derivative of the entanglement entropy across the phase transition.\footnote{One can check that the presence of the delta function is crucial for correctly computing the lengths of bulk curves via Crofton's formula as a volume in kinematic space~\cite{Czech:2015qta}.}

The metric can be mapped to hyperbolic coordinates in the two finite regions, using the redefinitions to hyperbolic time 	
\ali{
&\tau = -\log \tanh \frac{\pi \alpha \mathcal R}{\beta} \qquad \qquad \alpha < \alpha_{\rm c}~,\label{thypBH1}\\
&\tau  = \log \tanh \frac{\pi (\pi - \alpha)\mathcal R}{\beta} \qquad \ \alpha > \alpha_{\rm c}~.\label{thypBH2}
}
combined with the angular redefinition
\be \chi = \frac{2\pi \mathcal R}{\beta}\theta~.\ee

This covers two disjoint portions of the two hyperbolic patches,
\ali{
	ds^2 = \mathcal L^2 (-d\tau^2 + \sinh^2 \tau \, d\chi^2) \qquad  
	\left\{\begin{array}{ll}
	-\infty < \tau < \log \tanh \frac{\pi (\pi - \alpha_{\rm c})\mathcal R}{\beta} \\ 
	-\log \tanh \frac{\pi \alpha_{\rm c}\mathcal R}{\beta} < \tau < \infty
	\end{array}\right.
}	
which, after a $\theta \rightarrow \theta+\pi$ rotation, are glued together along their constant $\alpha=\alpha_{\rm c}$ boundaries, with a defect along the identification corresponding to the delta function in eq.~\eqref{KBTZbh}. The resulting kinematic space is depicted as embedded in dS$_2$ in figure \ref{BTZqSubregion}.

The sign of the time coordinate $\tau$ in eqs.~\eqref{thypBH1}-\eqref{thypBH2} is arbitrary as far as the metric is concerned. It determines the orientation of the geodesics, or equivalently, which half of the dS hyperboloid is covered. The minus sign is chosen to map future infinity $\tau = \infty$ to $\alpha=0$, or to cover part of the \emph{upper} half of the dS hyperboloid (as for the BTZ string). This portion of kinematic space, which consists of the non-winding connected Ryu-Takayanagi curves, forms a subregion of the upper hyperbolic patch of dS$_2$. In the second phase $\alpha \rightarrow \pi - \alpha$ and $\theta \rightarrow \theta \pm \pi$, and the sign of $\tau$ is also reversed to undo the change in orientation of the geodesic. This portion of kinematic space, consisting of the family of disconnected Ryu-Takayanagi curves past the phase transition, is mapped to a subregion of the hyperbolic patch in the \emph{lower back} half of the hyperboloid. 

Additional regions in the full hyperbolic patch of dS$_2$ that no longer belong to kinematic space correspond to geodesics that become winding upon quotienting, as illustrated in figure \ref{BTZBH1}.  
For a comparison of our picture to existing work on the kinematic space of the 2-sided quotiented BTZ black hole \cite{Czech:2015kbp}, see Appendix \ref{BTZAppendix}.

\subsection{Conical Singularity}\label{ConSing}

We can obtain a conical singularity geometry as a quotient AdS$_3/\mathbb{Z}_n$, where $\mathbb{Z}_n$ is a subgroup of the spatial rotation group $\mathrm{SO}(2)$ with $n$ an integer. Conical singularities with an arbitrary deficit angle are also solutions of 3d gravity, but we only consider $\mathbb{Z}_n$ singularities in this paper. We follow in this section the notation of~\cite{Balasubramanian:2014sra} and start from the metric of AdS$_3$,
\be ds^2 = -\left(1+\frac{R^2}{\lAdS^2}\right)dT^2 + \left(1+ \frac{R^2}{\lAdS^2}\right)^{-1} dR^2 + R^2 d\tilde \phi^2~, \ee
where $\lAdS$ is the AdS radius, but covering only the restricted angular range
\be -\frac{\pi}{n} \leq \tilde \phi \leq \frac{\pi}{n}~, \label{FDcon} \ee
with the endpoints identified.

\begin{figure}[t!]
	\centering	
	\includegraphics[width=4.5in]{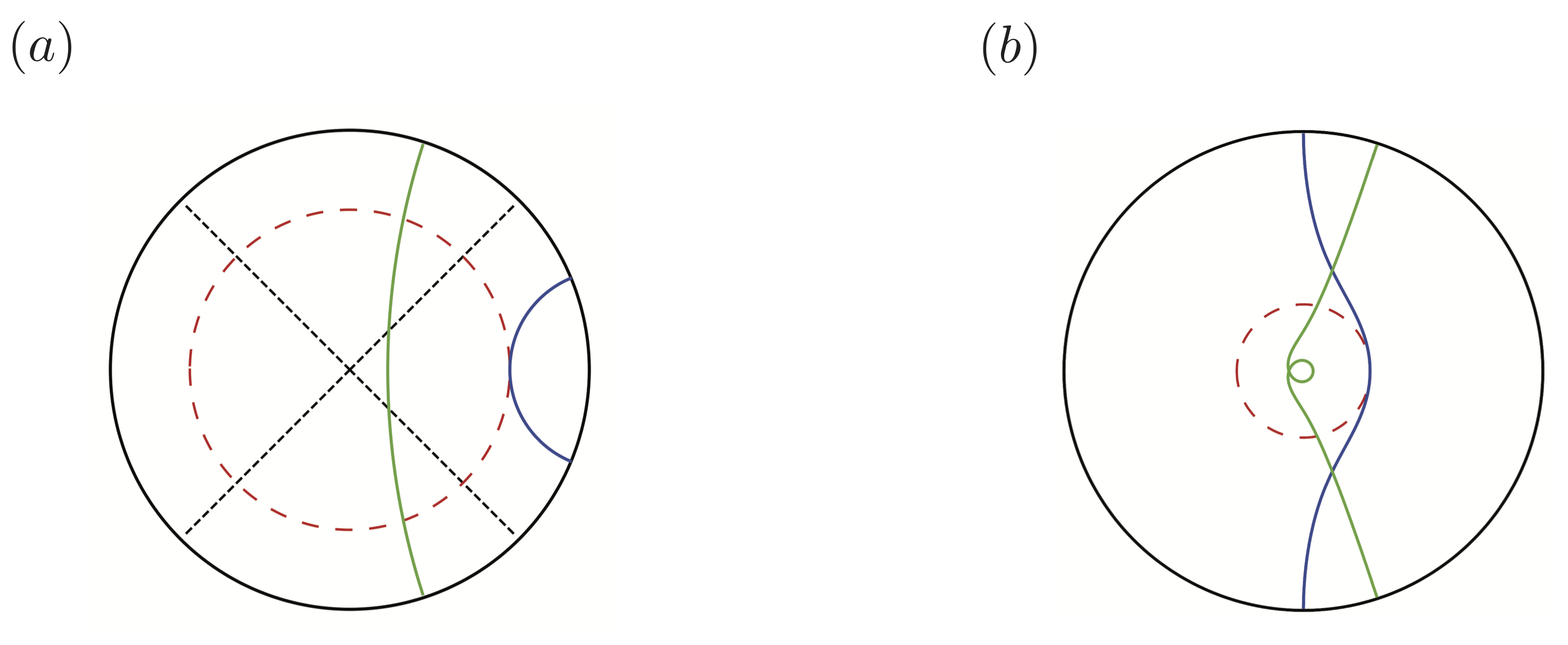} 
	\caption{(a) Two example geodesics on the covering AdS slice and (b) their counterparts on the slice of the quotiented AdS$/\mathbb{Z}_n$ geometry for $n=4$. The covering space consists of $n=4$ copies of the fundamental domain \eqref{FDcon}, marked by black dashed boundaries. The red dashed circles are the respective entanglement shadows $R_{\rm crit}(\tilde \alpha_{\rm max} = \pi/2n)$ and $r_{\rm crit}(\alpha_{\rm max} =\pi/2)$, defined in \eqref{rcritstar}. The blue geodesic has the maximum allowed opening angle, $\alpha = \pi/2$, in order to be minimal. 
		The green geodesic with $\alpha > \pi$ winds around the singularity once and enters the entanglement shadow region.}
	\label{congeodesics}
\end{figure} 

Alternatively, we can change coordinates to
\bea
\phi &=& n \tilde\phi~, \\
r &=& \frac{R}{n}~, \label{rtransf}\\
t &=& n T~.
\eea
This gives the standard metric for the conical singularity~\cite{Balasubramanian:2000rt, Balasubramanian:2005qu},
\be ds^2 = -\left(\frac{1}{n^2} + \frac{r^2}{\lAdS^2}\right) dt^2 + \left(\frac{1}{n^2}+\frac{r^2}{\lAdS^2}\right)^{-1} dr^2 + r^2 d\phi^2 \qquad (-\pi \leq \phi\leq \pi)~.\label{consingmetric}\ee 
The dual CFT lives on the conformal boundary of this metric, which we take to be the cylinder $\mathcal C(2\pi \mathcal R)$ with a compact space-like direction $\phi$ and an infinite time-like direction, following the discussion in Appendix \ref{Background}.

In this geometry, parent geodesics of AdS get mapped to geodesics that wind up to $(n-1)$ times around the singularity at the origin, depending on how many times they cross the fundamental domain of the quotient: $(k-1)$-winding geodesics ($\theta,\alpha$) have $-\pi \leq \theta \leq \pi$ and $(k-1) \pi \leq \alpha < k \pi$. See figure~\ref{congeodesics} for some example geodesics in the covering space and its quotient.

The quotiented geodesics descend from the solutions in pure AdS (with AdS radius $\lAdS$). Given a midpoint angle $\theta$ and opening angle $\alpha$ in the conical singularity geometry \eqref{consingmetric}, the radial profile of a geodesic is
\be 
\label{eq:geo_con}
r(\phi;\alpha,\theta) = \frac{\lAdS}{n} \frac{\cos{\frac{\alpha}{n}}}{\sqrt{ \cos^2{\frac{(\phi-\theta)}{n}}-\cos^2{\frac{\alpha}{n}}}} \qquad (\theta - \alpha \leq \phi \leq \theta + \alpha)~.\ee
For a given opening angle, the maximum radial extent of the geodesic is
\be r_{\rm crit}(\alpha) = \frac{\lAdS}{n} \cot{\frac{\alpha}{n}}~.\label{rcritalpha}\ee

We will compute the entanglement entropy of a dual CFT interval from the length of the corresponding geodesic, but first we need to carefully enforce the homology condition of holographic entanglement entropy. We distinguish two cases:\\

\noindent{\bf Star}: We can imagine replacing the singularity with a small star with negligible back-reaction on the geodesics. The outside geometry is the same, with the key difference that there is no boundary to spacetime at the origin. This allows complementary intervals to share Ryu-Takayanagi curves while satisfying the homology condition. This corresponds to considering a pure state on the boundary (any entropy carried by the star is taken to be subleading in $c$).  

Since geodesics can be effectively deformed through the origin, for a given interval $[u,v]$ there are now two homologous boundary-anchored geodesics. These consist of a geodesic that does not wrap around the origin with respect to the interval, as well as the analogous curve for its complement $\alpha \rightarrow \pi - \alpha$ and $\theta \rightarrow \theta \pm \pi$, but taken with the opposite orientation. The Ryu-Takayanagi curve is the one with minimal length, which is always the curve that does not wrap around the origin. There are two phases: for $\alpha < \pi/2$ the Ryu-Takayanagi curve is given by eq.~\eqref{eq:geo_con}, and for $\alpha > \pi/2$ it is equal to the corresponding curve for the complementary interval, with opposite orientation.

\begin{figure}[t!]
	\centering 
	\includegraphics[width=2.8in]{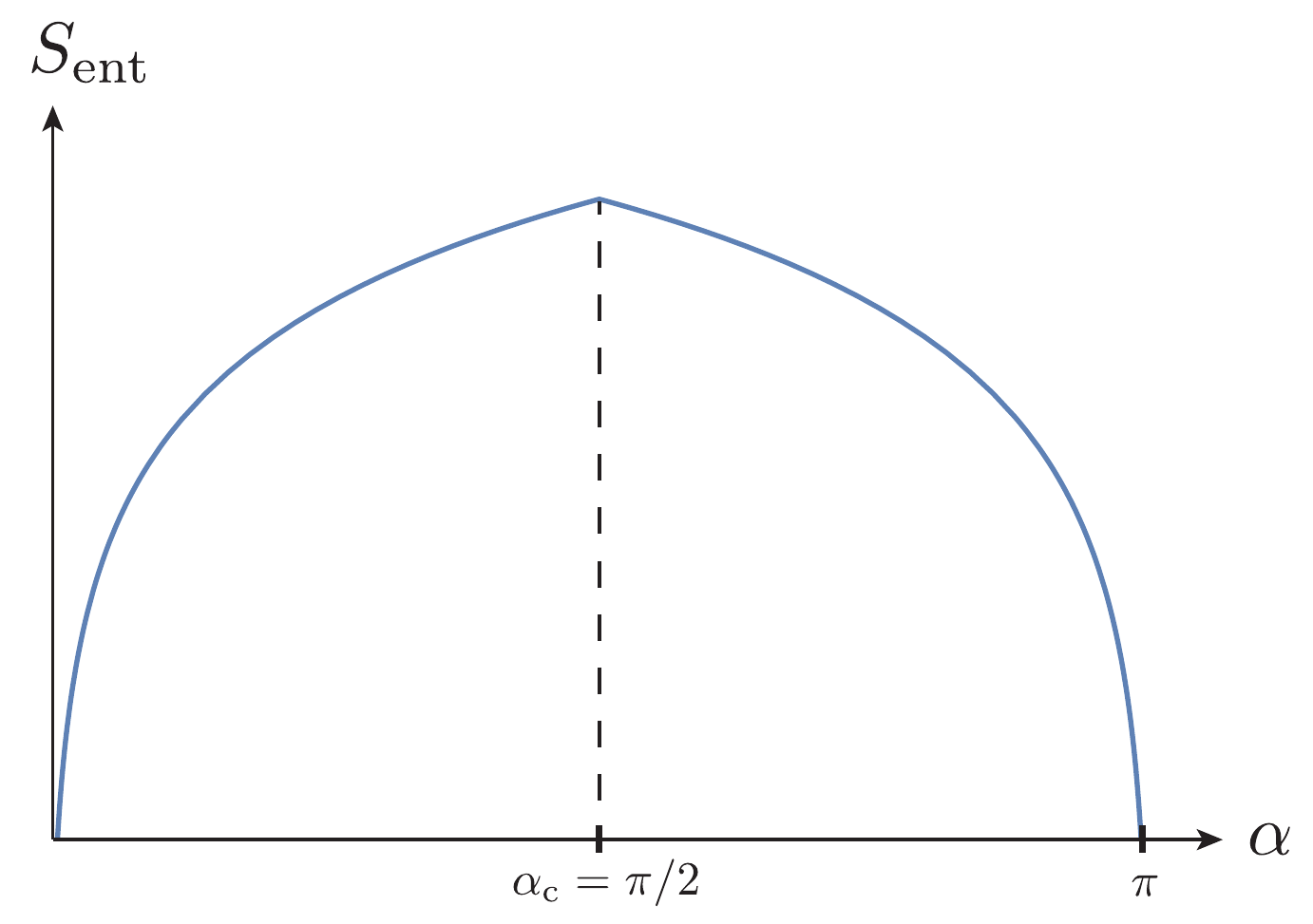} 
	\caption{The entanglement entropy of boundary intervals for the conical singularity, as a function of the opening angle $\alpha$ of the interval. At the critical angle $\alpha_{\rm c}=\pi/2$, there is a phase transition and a second family of complementary Ryu-Takayanagi geodesics have minimal length.}
	\label{Sent-star} 
\end{figure}

The entanglement entropies computed from these geodesics are given by (cf.~eq.~\eqref{eq:EEcirc})
\ali{
	\label{eq:EE_con}
	S_{\rm ent} = \left\{ \begin{array}{ll} \frac{c}{3} \log \left( \frac{2 n \mathcal R}{\epsilon} \sin \frac{\alpha}{n} \right) \qquad \ \ &\alpha < \frac{\pi}{2}~,  \\
		\frac{c}{3} \log \left( \frac{2 n \mathcal R}{\epsilon} \sin \frac{\pi-\alpha}{n} \right) \qquad &\alpha > \frac{\pi}{2}~, \end{array} \right. 
	}
shown in figure \ref{Sent-star}.	
The entanglement shadow is defined by the minimal radius probed by Ryu-Takayanagi geodesics: 
\be r_{\rm crit}(\alpha_{\rm max}) = \frac{\lAdS}{n} \cot{\frac{\pi}{2n}}~.\label{rcritstar}\ee 
It is shown in figure \ref{congeodesics} as a red dashed circle.
\begin{figure}[t!]
	\centering 
	\includegraphics[width=5in]{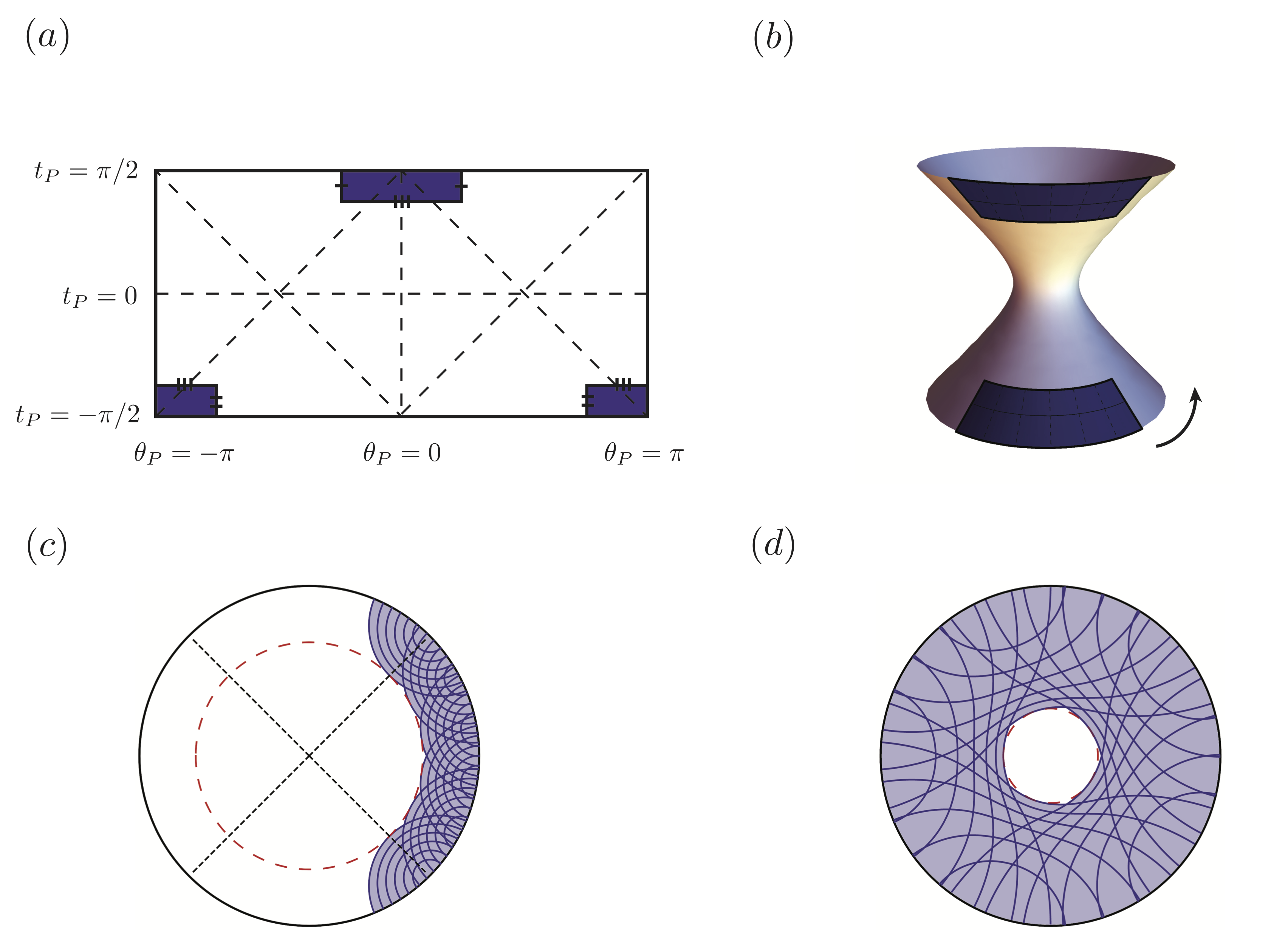} 	
	\caption{The kinematic space for the conical singularity covers two ``sub-de Sitter" patches, depicted in purple for the case $n=4$  in (a) the Penrose diagram, and (b) the embedding diagram, with lines of constant time (black) and constant $\theta$ coordinates (dashed). (The arrow in the embedding diagram indicates that the lower patch is actually located on the reverse side of the hyperboloid.) For the star, geodesics can be deformed through the origin, and the Ryu-Takayanagi geodesics satisfy $\alpha \leq \pi$. This results in two phases for the entanglement entropy and correspondingly, two distinct portions of kinematic space that, after a $\theta \rightarrow \theta + \pi$ rotation, are glued together on their constant $\alpha = \pi/2$ boundaries, with a defect placed along the identification. For a given $n$, the entire purple region takes up an $n$th fraction of the full de Sitter. In all cases the vertical boundaries of these sub-de Sitter patches are identified. (c) The region of the spatial slice of the covering AdS space covered by Ryu-Takayanagi geodesics. The maximum radial reach of these geodesics (eq.~\eqref{rcritstar}) delineates the entanglement shadow region around the origin (red dashed).  (d) The Ryu-Takayanagi geodesics cover the spatial slices of the conical singularity geometry up to the entanglement shadow. 
	}
	\label{CSSubregion}
\end{figure}

The kinematic space prescription in eq.~\eqref{refinedprescription} gives the metric
\be ds^2 = \frac{\ldS^2}{n^2} g(\alpha) (-d\alpha^2 + d\theta^2) \qquad (0 \leq \alpha \leq \pi, \ -\pi \leq \theta \leq \pi)~,\label{starmetric}\ee
where
\ali{
g(\alpha) = \left\{
\begin{array}{ll}
\csc^2{\left(\frac{\alpha}{n}\right)}  \qquad &0 \leq \alpha < \frac{\pi}{2} \nonumber\\
2n \cot\left({\frac{\pi}{2n}}\right)\, \delta\left(\alpha - \frac{\pi}{2}\right) \qquad &\alpha = \pi/2 \nonumber \\
\csc^2{\left(\frac{\pi-\alpha}{n}\right)}  \qquad &\frac{\pi}{2} < \alpha \leq \pi 
\end{array}\right.~.
}

\begin{figure}[h!]
	\centering
	\includegraphics[width=5in]{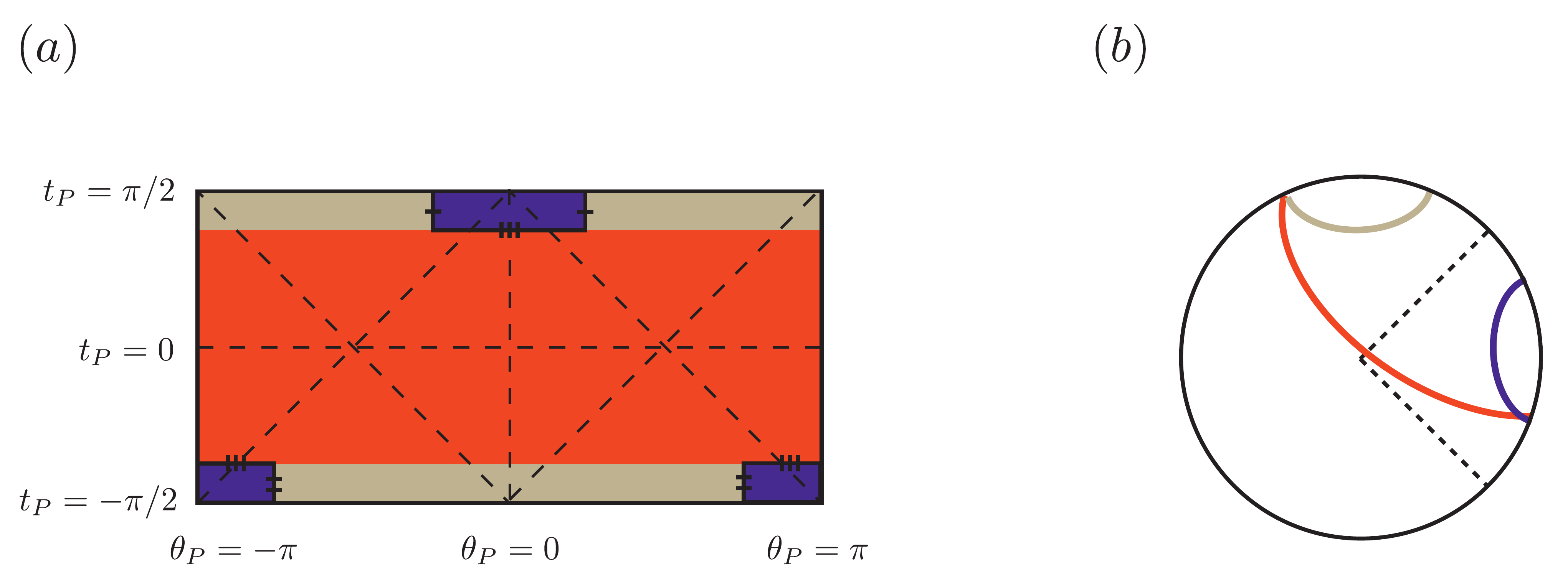}
	\caption{
		(a) The Penrose diagram for the kinematic space of the covering AdS space, color-coded to show the regions of geodesics that upon quotienting to the conical singularity geometry remain minimal (purple and brown) or become non-minimal, i.e., wrapping or winding (orange).  
		The innermost purple region is the kinematic space of the conical singularity geometry. Its constant $\theta$ boundaries are identified in the quotient with each other, along with each subsequent boundary of the $n$ fundamental domain copies $\theta \sim \theta + 2 \pi$ (in brown).  
		The constant-$\alpha$ lower boundary marks the maximum opening angle before minimal geodesics become non-minimal ones.    
		(b) An example geodesic for each region depicted on the Poincar\'e disk. 
		}\label{con-colors}	
\end{figure}

In the finite regions, we can convert to global coordinates through the angular transformation
\be \tilde \theta = \frac{\theta}{n}~, \ee
combined with the time redefinition
\begin{align}
\left\{\begin{array}{c}
t = -\log{\tan{\frac{\alpha}{2n}}} \qquad (0 \leq \alpha \leq \frac{\pi}{2})\\
t = \log{\tan{\frac{\pi-\alpha}{2n}}} \qquad \ (\frac{\pi}{2} \leq \alpha \leq \pi)
\end{array}\right.~.
\end{align}
The positive sign for the time coordinate has been chosen in the second phase to account for taking the opposite orientation of the geodesic associated to the complementary interval, since a flip in orientation maps to the opposite portion of the de Sitter hyperboloid. 

The metric maps to two subregions of global de Sitter space:
\begin{align} 
ds^2 = \ldS^2 \left(-dt^2 + \cosh^2 t d\tilde \theta^2\right)~ \quad
\begin{array}{c}
	\left\{\begin{array}{c} 
		-\infty \leq t < \log{\tan{\frac{\pi}{4n}}}\\
		-\log{\tan{\frac{\pi}{4n}}} < t < \infty
	\end{array}\right.~,
\end{array}
\end{align}
with $-\frac{\pi}{n} \leq \tilde \theta \leq \frac{\pi}{n}$ in all cases. 

From the embedding \eqref{globalembedding}, the kinematic space covers the two different regions of the global de Sitter depicted in figure~\ref{CSSubregion}, which we refer to as ``sub-de Sitters". The boundaries $\tilde \theta = -\frac{\pi}{n}, \frac{\pi}{n}$ correspond to geodesics that are identified by the quotient, and so the vertical boundaries are identified in kinematic space. The subregion connected to past infinity accounts for the geodesics past the phase transition, which are located on the bottom back half of the hyperboloid due to their reversed orientation. After a $\theta \rightarrow \theta + \pi$ rotation, the two sub-de Sitter regions corresponding to the two families of Ryu-Takayanagi geodesics (before and after the phase transition) are glued together along their $\alpha=\pi/2$ boundaries. Along the identification, there is a defect corresponding to the delta function contribution to the metric \eqref{starmetric}. The situation here is very similar to the effect of the phase transition for the BTZ black hole, see Section~\ref{BTZqSec}.

The regions in the full global dS$_2$ that no longer belong to kinematic space correspond to geodesics that become winding upon quotienting, illustrated in figure \ref{con-colors}. Such winding geodesics, with $\alpha > \pi$, are conjectured in~\cite{Balasubramanian:2014sra} to descend from a CFT concept called `entwinement', associated with entanglement between internal (gauged) degrees of freedom, rather than position space entanglement entropy. \\

\noindent {\bf True Conical Singularity}: For a true conical singularity geometry, the singularity is part of the boundary of spacetime, and the geodesics that compute the holographic entanglement entropy of intervals approaching the entire circle would be required to wrap around the singularity. In other words, the geodesics would have an opening angle in the range $0 \leq \alpha \leq \pi$. Since there is no horizon, it seems there would be no transition to a pair of disconnected geodesics (of the kind we saw for the BTZ black hole). However, for several reasons we are led to interpret this naked singularity as an unphysical idealization in the context of AdS$_3$/CFT$_2$ that should either be replaced by a smooth geometry or dressed in a possibly Planck-scale horizon. Either possibility results in the same behavior of the single-interval entanglement entropy and the same kinematic space, to leading order in $c$, that we discussed above for the star.

The first reason is that one generally expects the entanglement entropy of a subsystem to decrease as the subsystem approaches the total system. This comes from subadditivity applied to a system $A$ and its complement $\bar{A}$: $S_\mathrm{tot} = S_{A\cup \bar{A}} \leq S_A + S_{\bar{A}}$. Since $S_{\bar{A}} \to 0$ as $A$ approaches the entire system, we expect that $S_A$ will approach $S_\mathrm{tot}$ from above. Assuming no transition in the geodesics, i.e., using the entanglement entropy from only the first part of eq.~\eqref{eq:EE_con} in the range $0 \leq \alpha \leq \pi$, leads to a monotically increasing holographic entanglement entropy and one can check that this violates subadditivity for $\alpha$ approaching $\pi$. 

The second reason comes from CFT$_2$. In a number of recent works, e.g., \cite{Caputa:2014vaa, Asplund:2014coa}, asymptotically AdS$_3$ conical singularity geometries have been identified as dual to CFT$_2$ states excited by the insertion of a heavy primary operator. These CFT states are pure states, by construction, and are interpreted as dual to a geometries whose singularities are smoothed out by the presence of a star, as we studied above. One could construct a mixed state by combining many such pure states, but the maximal von Neumann entropy of such a mixed state is roughly $\log \Omega$, where $\Omega$ is the density of such states at a given conformal weight. We expect that this degeneracy is bounded by the Cardy formula \cite{Cardy:1986ie, Asplund:2015eha}
\begin{equation}
	\Omega(L_0, \bar{L}_0) \approx \exp\left(2\pi \sqrt{\frac{c}{6} L_0} + 2\pi \sqrt{\frac{c}{6} \bar{L}_0}\right),
\end{equation}
where we are considering cases where the conformal weights $L_0 = \bar{L}_0$ are large, corresponding to heavy states. 
This would lead to a maximal entropy that goes like $c^{1/2}$ in the large $c$ limit, which is subleading.
Hence, as far as holographic entanglement entropy is concerned, such states are effectively pure states.

Finally, recent work \cite{Casals:2016ioo, Emparan:2002px} indicates that adding a quantum field to the geometry and including backreaction dresses the singularity with a Planck-scale horizon. This picture leads one to expect a transition in the holographic entanglement entropy very similar to that discussed for the BTZ black hole, but where the horizon-wrapping geodesic is effectively reduced to a point at $r = 0$ and has zero length. The resulting single-interval entanglement entropy is then the same as we discussed above for the star geometry. \\

\section{Causal Structure}\label{Causal}

\begin{figure}[t!]
	\centering
	\includegraphics[width=5in]{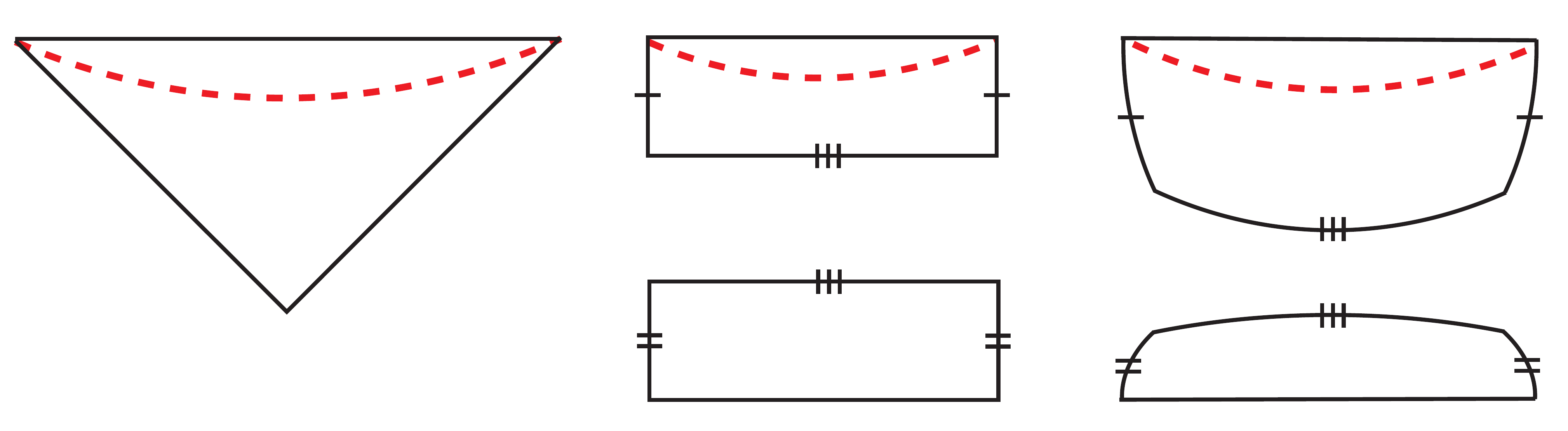}
	\caption{Different possible shapes for the Penrose diagram for kinematic space, for the Poincar\'e patch and BTZ black string (left), a conical singularity (center), and the BTZ black hole (right). For the conical singularity and the BTZ black hole, the time-like and space-like lines are separately identified. In all cases the resulting spacetimes are globally hyperbolic, with Cauchy surfaces whose domain of dependence is the full spacetime depicted as curved red lines.}
	\label{GlobalHyperbolicity}
\end{figure}

Up to this point, we have considered all the kinematic spaces we found as subsets of dS$_2$. In this section we want to consider these kinematic spaces as distinct spacetimes in their own right. In the examples we have considered, these spacetimes share an important causal property: they are globally hyperbolic and admit Cauchy surfaces at future infinity. To see this, the Penrose diagrams for the various kinematic spaces as well as sample Cauchy surfaces for each are schematically depicted in the case of the Poincar\'e patch, BTZ black string, BTZ black hole, and conical singularity in figure~\ref{GlobalHyperbolicity}.

For the quotiented BTZ black hole and conical singularity, the identification of geodesics along a boundary is crucial for the causal structure of kinematic space. Indeed, we have seen that the Penrose diagrams for these examples are unions of convex rectangular regions (with curved boundaries in the case of the BTZ black hole) that are  glued together in the case of phase transitions. Without the identification of the vertical boundaries, a light ray emanating from a point inside the subregion could end at these boundaries rather than the Cauchy surface and the spacetimes would not be globally hyperbolic.

It is also important to note that as \emph{subsets of dS$_2$}, these regions may form a \emph{proper subset} of the domain of dependence of any Cauchy slice. They form globally hyperbolic regions only once they are considered as distinct spacetimes with boundary.

We will now argue that this feature is a quite general property of kinematic space. We distinguish two cases, first for non-quotient space geometries (such as global AdS$_3$, the Poincar\'e patch, and the BTZ black string), and second for quotient space geometries (such as the BTZ black hole or conical singularity). \\

\noindent {\bf Non-quotient space case}: The various non-quotient space solutions of asympto\-tically AdS$_3$ gravity, restricted to a space-like slice, cover different regions of two-dimensio\-nal hyperbolic space and may be depicted as regions of the Poincar\'e disk or upper half plane (see Appendix~\ref{Background}). In all the cases we know of, these regions are either unbounded or are unions of fundamental domains, which we may assume are bounded by geodesics.

\begin{figure}[t!]
	\centering
	\includegraphics[width=6in]{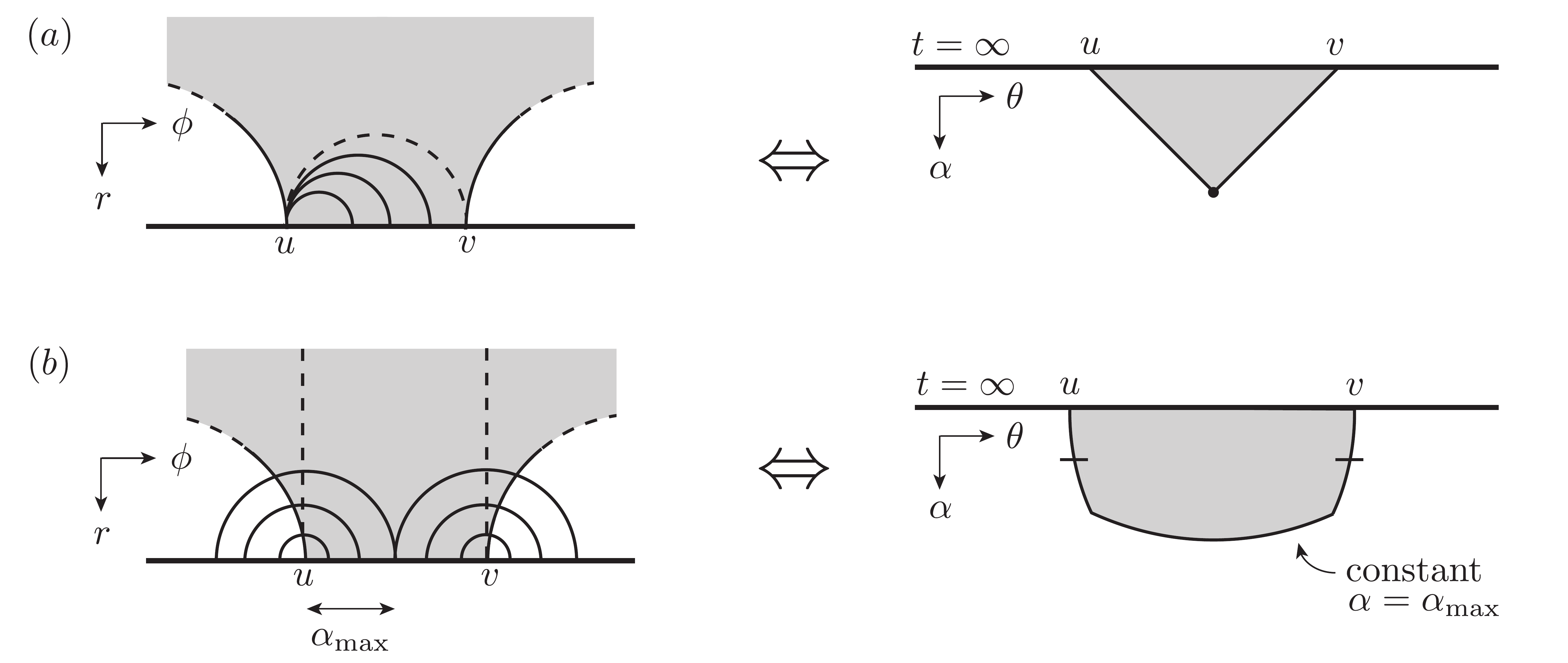}
	\caption{(a) To the left, an arbitrary subregion of the space-like slice that is bounded by geodesics (semi-circles and vertical lines in the hyperbolic plane) and intersects the boundary. Geodesics that lie fully inside this region map to a forward lightcone that ends at the future boundary in kinematic space (right). The kinematic boundaries are null because they correspond to geodesics that share an endpoint, with the largest contained geodesic (dashed) corresponding to the point at the tip of the lightcone. (b) If we instead consider a quotient whose fundamental domain is the shaded region to the left, geodesics that exit the region may still be represented in kinematic space. The time-like kinematic space domain boundaries consist of all geodesics with a midpoint at either endpoint of the boundary interval (represented as dashed vertical lines). Due to the quotient, these geodesics are identified. Any space-like boundary is a  constant-$\alpha$ line, which for example could correspond to a maximum opening angle.}
	\label{KSBoundary}
\end{figure}

The bulk regions will, in general, intersect the conformal boundary at a collection of disjoint intervals, each corresponding to a disjoint region in kinematic space.\footnote{Since kinematic space is currently formulated for single intervals only, we disregard geodesics that connect disjoint boundary regions, which would correspond to the entanglement entropy of multiple intervals. It would be interesting to generalize the proposal to include these cases.} For a given connected boundary interval, all geodesics confined to anchor to this interval are in the causal future of the maximal geodesic that connects the two endpoints of the interval. The kinematic space boundaries are null, corresponding to geodesics that share a single endpoint of the boundary interval, and these boundaries intersect at the point corresponding to the maximal geodesic. Thus, this portion maps to the filled-in forward light-cone in kinematic space that intersects the space-like future infinity of the ambient de Sitter space (see figure~\ref{KSBoundary}a). Such regions are manifestly globally hyperbolic, with the future boundary as a Cauchy surface.

Examples of non-quotient space geometries include the Poincar\'e patch and the BTZ black string, and both their kinematic spaces are future lightcones that intersect the future boundary (see figures \ref{PoincareSubregion} and \ref{BTZStringSubregion}).\\

\noindent {\bf Quotient space case}: Additional solutions can be obtained as quotients by a subgroup of the AdS$_3$ isometry group SO($2,2$). On the space-like slice, these subgroups descend to subgroups of the M\"{o}bius transformations. If we consider only discrete subgroups, known as Fuchsian groups, then the actions of the subgroups tesselate the hyperbolic disk or plane into polygonal fundamental domains with geodesic boundaries. These fundamental domains are identified under the quotient \cite{Aminneborg:1997pz, Brill:1998pr, dal2010geodesic, Balasubramanian:2014hda}.

In constructing kinematic space, a geodesic need not remain confined in a single fundamental domain, since the part of the geodesic that exits will be identified with a geodesic segment that is fully contained in the domain. The time-like domain boundaries of kinematic space consist of geodesics with constant $\theta$, i.e., geodesics whose midpoint is aligned with either boundary endpoint of the fundamental domain in the covering space (see figure~\ref{KSBoundary}b). The geodesics on either of these vertical domain boundaries of kinematic space are exactly identified under the quotient. Any remaining boundary is a (space-like) constant-$\alpha$ slice. For example, this could be $\mathcal{I}^+$ or $\mathcal{I}^{-}$ or it could correspond to a maximum opening angle. 

The quotient also introduces an additional subtlety due the possibility of phase transitions, when the geodesic length may be minimized by different classes of geodesics in different regions of parameter space. In this situation there is a critical $\alpha$ separating the contributions of each distinct family of geodesics. This leads to different patches of kinematic space covered by each type of geodesic, which are glued together along a constant time slice corresponding to this critical angle.

If the entanglement entropy or its derivative exhibits a kink across the phase transition, 
the metric may blow up along the glued interface (see for instance the delta function appearing in the metrics eqs.~\eqref{KBTZbh} and~\eqref{starmetric}). In the examples we considered, time-like and space-like geodesics crossing this defect still have finite length and the behavior of null geodesics indicates that the causal structure is not significantly affected. Additionally, we expect that these kinks are an artifact of setting $c=\infty$ and any divergences should be regulated when $1/c$ corrections are taken into account. Thus, we will assume that propagation through this defect is possible and well-defined. In this case, due to the identification of the constant $\theta$ lines in each region and the ability to propagate across the interface, kinematic space will still be globally hyperbolic.

Examples of quotient space geometries include the conical singularity and BTZ black hole\footnote{More topologically complex examples include the pair-of-pants wormholes considered in~\cite{Balasubramanian:2014hda}.}, which exhibit phase transitions (see figure~\ref{GlobalHyperbolicity}). In both cases the kinematic space is topologically a cylinder which is globally hyperbolic with a Cauchy surface at the future boundary. \\

The global hyperbolicity of these kinematic spaces implies that they are always causally well behaved, i.e., they can admit dynamical fields with well-posed initial value formulations \cite{Wald:1984rg}. Furthermore, the existence of Cauchy surfaces close to the future boundary means that boundary conditions set there determine the entire propagation within the interior.

\section{Relation to Auxiliary de Sitter}\label{AuxSpace}

The ``auxiliary de Sitter" proposal~\cite{deBoer:2015kda} provides a means for obtaining dynamics on an emergent de Sitter space of arbitrary dimension from the entanglement entropy of a conformal field theory. While reminiscent of kinematic space in two dimensions, the approach is less direct and not obviously equivalent:  rather than deriving the metric for a static spacetime, the authors observe that a de Sitter boundary-to-bulk propagator for a Klein-Gordon field is contained in the expression for the modular Hamiltonian, and that consequently perturbations of the entanglement entropy satisfy the de Sitter wave equation.

We begin by reviewing the original discussion in \cite{deBoer:2015kda} for a CFT in the vacuum on a plane, then proceed to several generalizations. In each case we provide a match to kinematic space.

\subsection{Vacuum on a Plane} \label{auxreview}

Consider a spherical region $A$ on a constant time slice in  a $d$-dimensional CFT in flat spacetime $R^{1,d-1}$ (with coordinates $t, \theta'_1, \cdots, \theta'_{d-1}$). The ball $A$ has radius $\alpha$ and center $\vec \theta$. Given a density matrix $\rho_{\rm tot}$ for the full system, the reduced density matrix is $\rho = \tr_{\bar A} \rho_{\rm tot}$. The entanglement entropy with the rest of the system is $S_{\rm ent} = -\tr(\rho \log \rho)$, and the modular Hamiltonian $H_{\rm mod}$ is defined by $\rho_A = e^{- H_{\rm mod}}/(\tr e^{- H_{\rm mod}})$. Given a CFT in its vacuum state, the modular Hamiltonian for $A$ can be derived by conformally mapping $A$ to the half-line, which has the Rindler wedge as its causal development region and consequently has a modular Hamiltonian that is the generator of Rindler time translations~\cite{Casini:2011kv}:
\ali{
 H_{\rm mod} = 2\pi \int_A d^{d-1} \vec \theta' \,\, \frac{\alpha^2 - (\vec \theta'-\vec \theta)^2}{2\alpha} T_{00}(\vec \theta')~, \label{Hmod}
}
where $\vec \theta^2 = \theta_1^2 + \cdots \theta_{d-1}^2$ and $T_{00}(\vec \theta')$ is the energy density operator.

\begin{figure}[t!]
	\centering
	\scalebox{1}{
		\includegraphics[width=2.5in]{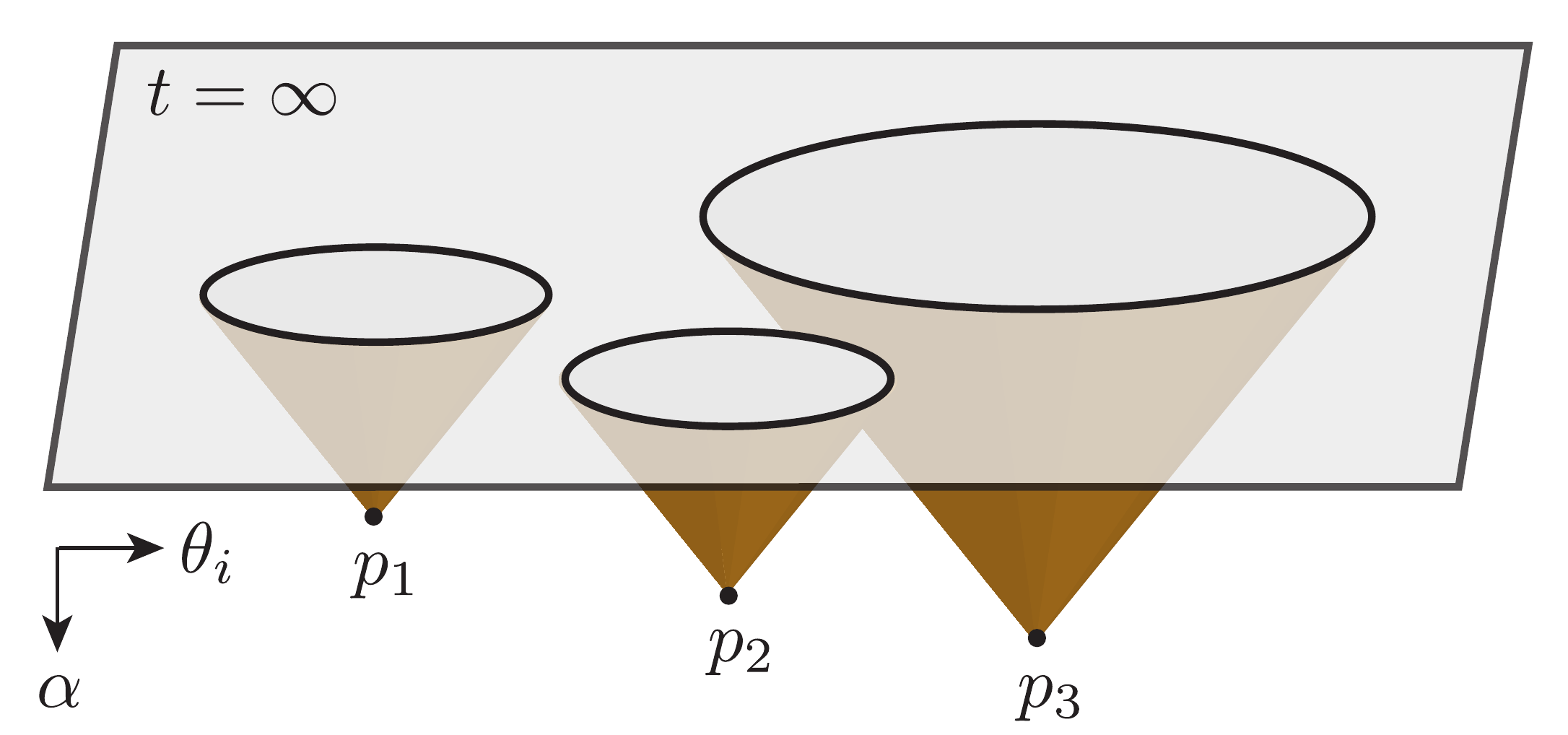} 
		}
	\caption{The mapping between points $p_i$ in auxiliary de Sitter dS$_d$ to spherical regions on the boundary CFT. The conformal boundary is identified with the asymptotic future of dS$_d$. A point $p_i$ is at the tip of the lightcone that projects to the corresponding ball.}
	\label{myersfig}
\end{figure}

It is observed in~\cite{deBoer:2015kda} that the fraction in the integrand of eq.~\eqref{Hmod} is a boundary-to-bulk propagator 
\be \mathcal P_{\rm planar} =  \frac{\alpha^2 - (\vec \theta'-\vec \theta)^2}{2\alpha}~ \label{planarprop}\ee  
of a scalar field of mass $m^2 = -d/\ldS^2$ on a $d$-dimensional de Sitter space in planar coordinates 
\be ds^2 = \frac{\ldS^2}{\alpha^2} (-d\alpha^2 + d\vec \theta^2)~\label{metricreview}~,\ee
with the de Sitter time coordinate given by the radius of the sphere. 
If $\delta S_{\rm ent}$ is the difference between the entanglement entropy of a slightly excited state and the entanglement entropy of the reference vacuum state, the ``first law of entanglement entropy'' tells us that\footnote{
		This can be derived from the relative entropy between an excited state and a reference state, defined as $S(\rho^{\rm exc}|\rho) = \tr(\rho^{\rm exc} \log \rho^{\rm exc}) - \tr(\rho^{\rm exc} \log \rho)$. Alternatively, the relative entropy can be written as $S(\rho^{\rm exc}|\rho) = \delta \langle H_{\rm mod} \rangle - \delta S_{\rm ent}$, with $\delta \langle H_{\rm mod} \rangle = \tr(\rho^{\rm exc} H_{\rm mod}) - \tr(\rho H_{\rm mod})$ and $\delta S_{\rm ent} = S_{\rm ent}(\rho^{\rm exc}) - S_{\rm ent}(\rho)$ the differences of the expectation value of the modular Hamiltonian and the entanglement entropy with respect to the reference state.   Relative entropy has the property that it is always positive, and in the limiting case of infinitesimally small excitations, this leads to the first law of entanglement.}
$\delta S_{\rm ent} = \delta \left< H_{\rm mod}\right>$.  
By eq.~\eqref{Hmod}, combined with the first law of entanglement entropy, the perturbation $\delta S_{\rm ent}(\alpha, \theta)$ is a scalar field that solves the de Sitter wave equation in planar coordinates, with future ($\alpha \to 0$) boundary conditions set by $\langle T_{00}(\theta') \rangle$. That is,
\be(\nabla^2 - m^2)\delta S_{\rm ent} = 0~, \quad \text{with $\quad m^2 = -\frac{d}{\ldS^2}$}~. \label{KGeq} \ee

The authors argue for a visualization of the mapping between the CFT and the emergent de Sitter space as follows. The constant time slice of the CFT is taken to be the future  asymptotic boundary $\mathcal I_+$ of dS$_d$. Each point in de Sitter corresponds to a ball in this time slice via the intersection of its future lightcone with $\mathcal I_+$ along the boundary of the ball (see figure~\ref{myersfig}). The causal structure of dS$_d$ directly translates into containment relations between spherical regions: a ball $A$ contained in a ball $B$ is said to be in the time-like future of $B$. In $2$ dimensions, the balls are intervals and this replicates the causal structure of kinematic space discussed in Section~\ref{KinSpace}. Indeed, the mapping in figure~\ref{myersfig} reduces to the mapping in figure~\ref{Embedding}a.   
\\

\noindent{\bf Matching to kinematic space of the Poincar\'e patch of AdS: }  For a CFT in the vacuum on a plane $R^{1,1}$, the auxiliary dS metric \eqref{metricreview} is the same as the kinematic space metric for the Poincar\'e patch of AdS~\eqref{Kplanar}. 
The region of de Sitter that is covered is the planar patch depicted in figure~\ref{PoincareSubregion}.

\subsection{Vacuum on a Cylinder} \label{auxvaccyl}

Consider now the case of a vacuum CFT on a cylinder $R^1 \times S^1$ with radius $\mathcal R$. The spatial, angular coordinate $\theta'$ measures the distance along the periodic space-like dimension. We consider an interval of angular extent $2 \alpha$, or length $L = 2 \mathcal R \alpha$, centered around $\theta$, on a constant time slice.  

The modular Hamiltonian for the interval can be obtained by applying a conformal transformation to the planar result~\eqref{Hmod}, and is given by~\cite{Blanco:2013joa, Casini:2011kv}
\be H_{\rm mod} = 2\pi \mathcal R^2 \int_{\theta-\alpha}^{\theta+\alpha}d\theta' \frac{\cos(\theta'-\theta) - \cos \alpha}{\sin \alpha} T_{00}(\theta')~.\label{globalmod}\ee

We could again associate with this modular Hamiltonian an emergent auxiliary de Sitter space if the fraction in the integrand of \eqref{globalmod} can be recognized as a boundary-to-bulk propagator. Motivated by the equivalence of the auxiliary de Sitter and kinematic space for the Poincar\'e patch of AdS, we make the ansatz that the auxiliary de Sitter associated with eq.~\eqref{globalmod} is the kinematic space of global AdS$_3$, given in eq.~\eqref{globalcoords}:  
\be ds^2 = \frac{\ldS^2}{\sin^2\alpha} (-d\alpha^2 + d\theta^2) 
~. \label{globalmetric} \ee

To write down the expression for the boundary-to-bulk propagator on this global de Sitter space, it is easiest to write the propagator on the planar patch (given by eq.~\eqref{planarprop}) in embedding coordinates first (using \eqref{planarembedding}), 
\be \mathcal P = \frac{X-U}{2}~. \label{propemb} \ee 
This can be subsequently transformed to global coordinates (using \eqref{globalembedding} and \eqref{tOfalphaglobal}), which results in
\be \mathcal P_{\rm global} = \frac{\ldS}{2} \frac{\cos(\theta'-\theta) - \cos \alpha}{\sin \alpha}~.\label{globalprop}\ee 

We can ``normalize" the propagator $\mathcal P_{\rm global}$ to have the same behavior as $\mathcal P_{\rm planar}$ near the limits of the interval:
\bea
\mathcal P_{\rm planar} &=& - (\theta'-\theta-\alpha) + \mathcal O\left((\theta'-\theta-\alpha)^2\right)~, \qquad \quad \ \ \ \theta'\rightarrow \theta +  \alpha~,\label{planarlimit}\\
\mathcal P_{\rm global} &=& -\frac{\ldS}{2}(\theta'-\theta - \alpha) + \mathcal O\left((\theta'-\theta - \alpha)^2\right)~, \qquad \quad \theta' \rightarrow \theta + \alpha~,  \label{globallimit}
\eea
requiring us to set\footnote{To see this, we refer to the comment on notation on page \pageref{notationcomment} to note that $\theta$ in \eqref{planarlimit} is a length while $\theta$ in \eqref{globallimit} is an angle.} 
\be \ldS = 2\mathcal R~.~\label{LR}\ee 
\\

\noindent{\bf Matching to kinematic space of global AdS: }  
From~\eqref{globalmod}, \eqref{globalprop} and \eqref{LR} we see that the modular Hamiltonian indeed takes the form 
\be H_{\rm mod} = 2\pi \int \mathcal P_{\rm global} \, T_{00}~. \label{globalHprop}\ee
This confirms our ansatz of identifying the auxiliary dS metric for a vacuum CFT on a cylinder with the kinematic space of global AdS~\eqref{globalcoords}, covering the full dS$_2$ depicted in figure~\ref{AdSSubregion}. The matching required fixing the de Sitter radius $\mathcal L$ to the circumference of the cylinder over $\pi$ in eq.~\eqref{LR}. 

By the first law of entanglement, the perturbation $\delta S_{\rm ent}$ of the entanglement entropy of the interval associated with small excitations with respect to the reference vacuum state is given by 
\be \delta S_{\rm ent} = \delta \left< H_{\rm mod} \right>
             = 2\pi \mathcal R^2 \int_{\theta-\alpha}^{\theta+\alpha}d\theta' \frac{\cos(\theta'-\theta) - \cos \alpha}{\sin \alpha} \left< T_{00}(\theta') - T_{00}^{\rm vac, cyl} \right> ~,\label{}\ee
with $\langle T_{00}^{\rm vac, cyl}\rangle = - \frac{c}{24 \pi \mathcal R^2}$~\cite{DiFrancesco1997}. 
This $\delta S_{\rm ent}$ solves the de Sitter wave equation \eqref{KGeq} in global coordinates, and as such defines a local dynamic degree of freedom on the kinematic space \eqref{globalcoords}. The corresponding mass is now fixed, as a consequence of fixing the de Sitter radius, to 
\be  m^2 = -\frac{2}{\ldS^2} = -\frac{1}{2 \mathcal R^2} = \frac{12\pi}{c} \langle T_{00}^{\rm vac, cyl}\rangle~. \label{}\ee

\subsection{Thermal State} \label{auxthermal}

We consider in this section a CFT in a thermal state on a cylinder $R^1 \times S^1$ with radius $\beta/\pi$. That is, the imaginary-time-like dimension  is compact with periodicity $\beta$, the inverse temperature of the state. The spatial coordinate $\theta'$ measures the distance along the space-like dimension. We consider an interval of length $L = 2 \alpha$ centered around $\theta$, on a constant time slice.

The modular Hamiltonian for the interval can be obtained by conformally mapping the interval to the half-line, for which the modular Hamiltonian is just the boost generator on the Rindler wedge, and is given by\footnote{
	The relation between our spatial coordinate $\theta'$ and the spatial coordinate $x$ in \cite{Hartman:2015apr} is $x = \theta' - \theta + \alpha = \theta'-u$. 
	} \cite{Hartman:2015apr} 
\be H_{\rm mod} = 2\beta \int_{\theta-\alpha}^{\theta+\alpha} d\theta' \ \frac{\sinh{\frac{\pi (\theta'-\theta + \alpha)}{\beta}} \sinh{\frac{\pi (\alpha - \theta'+\theta)}{\beta}}}{\sinh{\frac{2 \pi \alpha}{\beta}}} T_{00}(\theta')~. \label{thermalmod}\ee
An equivalent expression that is more similar in form to eq.~\eqref{globalmod} is 
\be H_{\rm mod} =  2\beta \int_{\theta-\alpha}^{\theta+\alpha}d\theta' \ \frac{\cosh \frac{2\pi\alpha}{\beta} - \cosh \frac{2\pi (\theta'-\theta)}{\beta} }{ 2 \sinh \frac{2 \pi\alpha}{\beta}} T_{00}(\theta')~.\label{thermalmod2}\ee

We could again associate with this modular Hamiltonian an emergent auxiliary de Sitter space if the fraction in the integrand of \eqref{thermalmod2} can be recognized as a boundary-to-bulk propagator. Motivated by the discussed equivalences of the auxiliary de Sitter spaces and kinematic spaces for AdS, we make the ansatz that the auxiliary de Sitter associated with eq.~\eqref{thermalmod2} is the kinematic space of the BTZ black string, given in eq.~\eqref{KBTZstring}:  
\be ds^2 = \frac{4\pi^2 \ldS^2}{\beta^2} \frac{-d\alpha^2 + d\theta^2}{\sinh^2{\left(\frac{2 \pi \alpha}{\beta}\right)}}~. \label{hypmodmetric} \ee 
This metric was identified in Section~\ref{BTZSec} as the hyperbolic patch of dS$_2$. 

To write down the expression for the boundary-to-bulk propagator on the hyperbolic patch of de Sitter, we use the expression \eqref{propemb} of the de Sitter propagator in embedding coordinates and eqs.~\eqref{hyperbolicembedding} and \eqref{tOfalphahyp} to express it in hyperbolic coordinates: 
\be \mathcal P_{\rm thermal} = \frac{\ldS}{2} \frac{\cosh{\frac{2\pi \alpha}{\beta}} - \cosh{\frac{2\pi (\theta'-\theta)}{\beta}}}{\sinh{\frac{2 \pi \alpha}{\beta}}}~.\label{thermalprop}\ee

We can again ``normalize" the propagator $\mathcal P_{\rm thermal}$ by considering the limit close to the boundary of the interval, 
\be \mathcal P_{\rm thermal} = -\ldS \frac{\pi}{\beta}(\theta'-\theta - \alpha) + \mathcal O\left((\theta'-\theta - \alpha)^2\right)~, \qquad \theta' \rightarrow \theta+ \alpha~, \ee
and comparing it to the behavior \eqref{planarlimit} of the planar propagator. For those to match, we fix the kinematic space de Sitter radius to 
\be \ldS = \frac{\beta}{\pi}~.\label{Lbeta}\ee \\

\noindent{\bf Matching to kinematic space of the BTZ black string: }  
From~\eqref{thermalmod}, \eqref{thermalprop} and \eqref{Lbeta} we see that the modular Hamiltonian indeed takes the form
\be H_{\rm mod} = 2\pi \int \mathcal P_{\rm thermal} \, T_{00}~. \label{thermalHprop}\ee
This confirms our ansatz of identifying the auxiliary dS metric for a thermal CFT on a cylinder with the kinematic space of the BTZ black string~\eqref{KBTZstring}, covering the hyperbolic patch of dS$_2$ depicted in figure~\ref{BTZStringSubregion}. The matching required fixing the de Sitter radius $\mathcal L$ to the circumference of the cylinder over $\pi$ in eq.~\eqref{Lbeta}.

By the first law of entanglement, the perturbation $\delta S_{\rm ent}$ of the entanglement entropy of the interval associated with small excitations with respect to the reference thermal state is given by 
\be \delta S_{\rm ent} = \delta \left< H_{\rm mod} \right> 
=  2\beta \int_{\theta-\alpha}^{\theta + \alpha}d\theta' \ \frac{\cosh \frac{2\pi\alpha}{\beta} - \cosh \frac{2\pi (\theta'-\theta)}{\beta} }{ 2 \sinh \frac{2 \pi\alpha}{\beta}} \left< T_{00}(\theta')- T_{00}^{\rm thermal} \right> ~,\label{}\ee
with $\langle T_{00}^{\rm thermal}\rangle = \frac{\pi c}{6 \beta^2}$ \cite{DiFrancesco1997}. 
This $\delta S_{\rm ent}$ solves the de Sitter wave equation \eqref{KGeq} in hyperbolic coordinates, 
and as such defines a local dynamic degree of freedom on the kinematic space \eqref{KBTZstring}. The corresponding mass is again fixed, as a consequence of fixing the de Sitter radius, to 
\be  m^2 = -\frac{2}{\ldS^2} = -\frac{2 \pi^2}{\beta^2} = -\frac{12\pi}{c} \langle T_{00}^{\rm thermal}\rangle~. \label{}\ee

\section{Discussion and Further Directions}  \label{discussion}

\subsection{Refined Prescriptions} \label{prescriptions}

We have shown that the two distinct prescriptions from \cite{Czech:2015qta, Czech:2015kbp} and \cite{deBoer:2015kda} for an emergent de Sitter space give equivalent results in the case of a CFT on a plane, on a cylinder with a compact spatial direction or on a cylinder with a compact imaginary time direction. The latter two cases are holographically dual to global AdS or the BTZ black string, respectively. 
Based on these examples we (slightly) reformulate the prescriptions for the cylinder cases, so that they lead to the same emergent dS geometry. In particular, we include a specification of the de Sitter radius $\ldS$ that depends on the length of the compact direction in the CFT. (For the CFT on a plane there is no length scale present. Correspondingly, $\mathcal P_{\rm planar}$ in \eqref{planarprop} does not depend on $\ldS$.) 

Consider a (1+1)-dimensional CFT on a cylindrical conformal boundary, dual to either (2+1)-dimensional global AdS or the BTZ black string. 
Each interval $A$ at a constant time in the CFT has a reduced density operator $\rho_A$ and a modular Hamiltonian $H_{\mathrm{mod}}$, defined by $\rho_A = e^{-H_{\rm mod}}/(\tr e^{-H_{\rm mod}})$. As we have seen, the modular Hamiltonian can be written as an integral, $H_{\rm mod} = 2\pi \int_A \mathcal P \ T_{00}$, and the integrand defines a boundary-to-bulk propagator $\mathcal P$ of a scalar field with mass $m^2 = -2/\mathcal L^2$ on an emergent de Sitter space with radius $\mathcal L = \mathcal S/\pi$. Here $\mathcal S$ is the circumference of the cylinder's compact dimension. We remark that with this choice for $\mathcal L$, the mass squared of the scalar is proportional to the energy density of the CFT with proportionality factor equal to $\frac{12 \pi}{c}$, times $-1$ when the compact coordinate is imaginary time. 

The emergent de Sitter space associated with the conformal boundary can then be identified with the kinematic space or space of Ryu-Takayanagi geodesics on the constant time slice of the bulk geometry, obtained through the prescription in \eqref{refinedprescription}: 
\ali{
	ds^2_{\text{$\mathcal K$ of CFT on cylinder}} &= \frac{12}{c} \ldS^2 \frac{\p^2 S_{\rm ent}(u,v)}{\p u \p v} du dv \qquad \text{with $\ldS = \mathcal S/\pi$}~. 
}

By equating these two emergent de Sitter geometries, we obtain a dynamical scalar field moving on the kinematic space. The authors of \cite{Czech:2015qta, Czech:2015kbp} have argued that the MERA tensor network is a discretization of kinematic space. Our results thus offer a potentially interesting new ingredient in the study of this MERA--kinematic space connection. 

We can also reverse the argument: knowing the entanglement entropies of closed intervals allows you to calculate the kinematic space of the constant time slice in the bulk. The boundary-to-bulk propagator $\mathcal P$ of a scalar field with mass $m^2 \sim T_{00}$ on the kinematic space can then be used to write down an expression for the modular Hamiltonian:  
\ali{ H_{\rm mod} = 2\pi \int_A \mathcal P \ T_{00}~.}
For global AdS or the BTZ black string, we were able to check this line of reasoning with known results, but it would be interesting to go beyond this.

\subsection{Beyond Universality}

The refined prescription in the previous subsection applies specifically to cases where the CFT lives on a cylinder that was obtained from a conformal mapping of the plane with no operator insertions. Does the equivalence extend beyond this?

The partition function of a CFT on a spacetime with a genus higher than zero is not universal, i.e., it depends on the full spectrum of operators of the CFT and not just on its central charge. Consequently, the entanglement entropies of intervals in states defined on such spacetimes are also not universal (see, e.g., \cite{2009JPhA...42X4005C}). The same is true for the entanglement entropies in generic excited states. We expect that the modular Hamiltonians of single intervals are similarly non-universal in these cases, and they may be non-local as well. We explore the two emergent de Sitter space prescriptions in two such cases below, and though we do not find precise matches, as we did in the cases in Section~\ref{AuxSpace}, we find some suggestive results and avenues for further investigation.

In the realm of holographic two-dimensional CFTs, we can implicitly define a CFT state (to leading order in $c$) by a bulk geometry. In Section~\ref{ConSing} we calculated the kinematic space of a conical singularity spacetime and in Section~\ref{BTZqSec} the kinematic space of the BTZ black hole. The latter is dual to a CFT on a spatial circle at finite temperature, which corresponds to a spacetime with the topology of a torus. The conical singularity is dual to a CFT on a spatial circle excited by the insertion of a heavy primary operator \cite{Asplund:2014coa}, or perhaps a statistical mixture of such states. 

The kinematic space we found for the conical singularity, with metric given by eq.~\eqref{starmetric}, 
suggests that for holographic CFTs the modular Hamiltonian of sufficiently small intervals in such a state is of the form:
\begin{align}
\label{eq:CS_Hmod}
H_{\mathrm{mod}}^{\mathrm{con}} =  2\pi n \mathcal R^2 \int_{\theta -\alpha}^{\theta + \alpha} d\theta'  \frac{\cos \frac{\theta'-\theta}{n} - \cos \frac{\alpha}{n}}{\sin \frac{\alpha}{n}} T_{00}^{\mathrm{con}}~,
\end{align}
where the length of the interval is $2\alpha$, the radius of the spatial circle is $\mathcal{R}$ and $\theta, \theta' \in [-\pi,\pi]$, and we are considering a quotient by a $\mathbb{Z}_n$ subgroup of the spatial rotation group $\mathrm{SO}(2)$. 
Because of the phase transition in the entanglement entropy, we would only expect this to hold for intervals with $\alpha < \pi/2$. 
The fraction in the integrand has the form of a boundary-to-bulk propagator for a scalar field on the sub-de Sitter kinematic space (see figure \ref{CSSubregion}). If we take the mass of the field to obey $m^2 = -2/\mathcal{L}^2$, as in \cite{deBoer:2015kda}, and if we normalize by examining the behavior of the propagator near the end of the interval ($\theta' \to \theta + \alpha$), in analogy to eq.~\eqref{planarlimit}, we find $\mathcal{L} = 2n\mathcal{R}$ and $m^2 = -1/2n^2\mathcal{R}^2$.

We do not know of a CFT calculation of the modular Hamiltonian of an interval in a conical singularity state. However, the R\'enyi entropies for such states are known in certain limits \cite{Asplund:2014coa, Caputa:2014vaa}. It would be interesting to calculate the spectrum of eigenvalues of the reduced density operator (the entanglement spectrum) from these results, using the techniques of \cite{2008PhRvA..78c2329C, Hung:2011nu}, and use this to check or modify eq.~\eqref{eq:CS_Hmod}.

Similarly, one could write down the boundary-to-bulk propagator on the kinematic space of the BTZ black hole \eqref{KBTZbh}. This would suggest an expression for the modular Hamiltonian of an interval on a spatial circle at finite temperature (of a holographic CFT). For sufficiently small intervals, it is just given by eq.~\eqref{thermalmod2} with the angles replaced by $\mathcal R$ times the angles.  For larger intervals, we would have to determine the effects of the phase transition in the Ryu-Takayanagi curves. As for the conical singularity case, we do not know of a CFT calculation of the modular Hamiltonians. However, the R\'enyi entropies have been extensively studied and are known in a variety of limits \cite{Barrella:2013wja, Datta:2013hba, Perlmutter:2013paa, Chen:2014hta, Chen:2015kua, Chen:2016lbu}, and one could again check the consistency of the entanglement spectra. 

How generally can one go from the kinematic space, which is relatively easy to calculate, to the modular Hamiltonian? 
The general procedure would be to go from the boundary-to-bulk propagator $\mathcal P$ of a scalar field on the kinematic space to the modular Hamiltonian of some region $A$ through the formula $H_{\mathrm{mod}} = 2\pi \int_{A} \mathcal P \, T_{00}$,  
where $T_{00}$ is the $00$ component of the energy-momentum tensor operator in the given CFT state 
(this can be determined from the asymptotic behavior of the bulk metric, in holographic cases).  
In this sense, the modular Hamiltonian would be obtained from the entanglement entropy $S_{\rm ent}$ through a loop that includes the kinematic space $\mathcal{K}$ and the auxiliary de Sitter space:
\ali{
	\begin{matrix} S_{\rm ent} & \longrightarrow & \mathcal K \\ 
		&                 & \biggr\downarrow \\ 
		H_{\mathrm{mod}} &  \longleftarrow & \text{auxiliary dS} \end{matrix} \qquad.
}
This would be remarkable since the modular Hamiltonian of an interval is equivalent to its density operator, which, a priori, has much more information than just the entanglement entropy.
In the cases considered in Section~\ref{AuxSpace}, the modular Hamiltonians were already known from CFT calculations, so the sketched loop served as a check rather than a prediction. We expect this to work only for certain CFTs in certain states, but this might include holographic CFTs in many states dual to classical bulk geometries.
This is consistent with recent work \cite{Jafferis:2015del}, which allows one to compute certain boundary modular Hamiltonians from bulk data using relative entropy.
This would be interesting to investigate further, along with the relationship between kinematic space and the auxiliary de Sitter space in higher dimensions and in time-dependent states.

\appendix

\section{$3$d Gravity}\label{Background}
AdS$_3$ is defined as the locus 
\equ{
	-U^2 - V^2 + X^2 + Y^2 =  -\lAdS^2 \label{AdSlocus}
}
in the flat embedding space $R^{2,2}$, with $\lAdS$ the AdS radius. The induced metric is
\equ{
	ds^2 = -dU^2 - dV^2 + dX^2 + dY^2~.  \label{metric}
	}
The locus \eqref{AdSlocus} is left invariant by $R^{2,2}$ Lorentz transformations $SO(2,2)$, the isometry group of AdS. 
Classifying these isometries into orbits in space-like and time-like planes, coordinate systems can be introduced that make different classes of isometries manifest. The resulting AdS$_3$ metrics correspond to different classes of solutions of classical, pure AdS$_3$ gravity. Indeed, $(2+1)$-dimensional classical gravity with a negative cosmological constant (and no source terms) is trivial in the sense that the solutions have constant negative curvature: 
all solutions are locally AdS$_3$ everywhere. Globally distinct solutions such as the BTZ black hole and the conical geometry are obtained as quotients of AdS$_3$.

\begin{figure}[t!]
	\centering
	\includegraphics[width=4in]{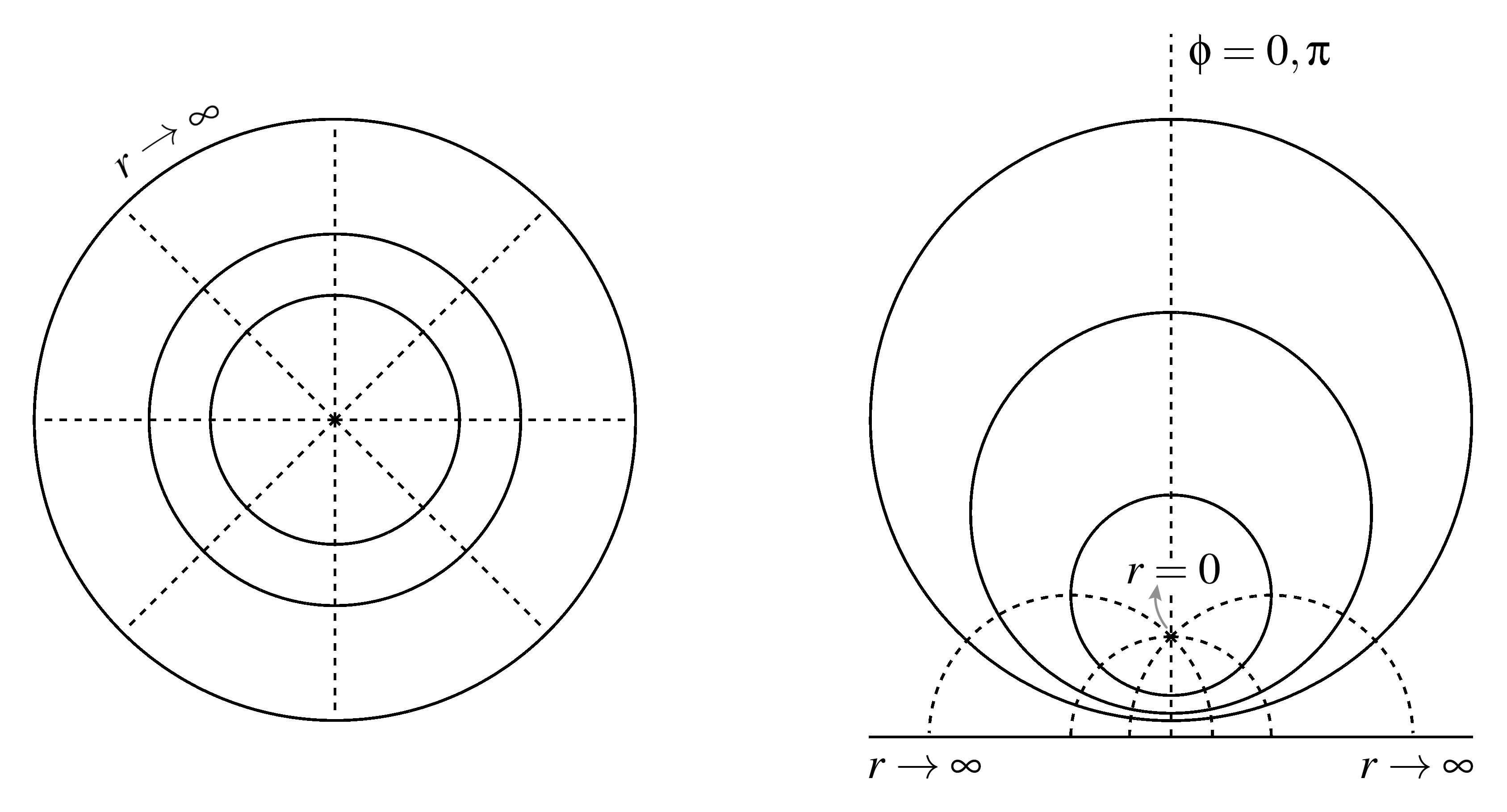}
	\caption{Constant time slice $H_2$ of AdS$_3$, represented as a Poincar\'e disk on the left and the half-plane on the right, showing constant $r$ (in solid black) and constant $\phi$ lines (black dashed). We can imagine mapping the boundary $r\rightarrow \infty$ to a circle of radius $\mathcal R$ instead of the unit radius Poincare disk, i.e. $y_i \rightarrow \mathcal R y_i$ in \eqref{Poincarecoord}. The radius of the disk is then $\mathcal R$, consistent with \eqref{confbdy}.
}
	\label{PoincareDiskAdS}
\end{figure}

In `static AdS coordinates'
\equ{
	\left( \begin{matrix} U \\ V \\ X \\ Y \end{matrix} \right)
     =  \left( \begin{matrix} \sqrt{r^2+\lAdS^2} \cos(t/\lAdS) \\ \sqrt{r^2+\lAdS^2} \sin(t/\lAdS) \\ r \cos \phi \\ r \sin \phi  \end{matrix} \right) \label{staticembcoord}
}
the metric \eqref{metric} takes the pure AdS$_3$ form:
\ali{
	ds^2	&= -\left(\frac{r^2}{\lAdS^2}+1 \right) dt^2 + \left(\frac{r^2}{\lAdS^2}+1 \right)^{-1} dr^2 + r^2 d\phi^2 \qquad (\text{AdS})~,    \label{AdSrthetat}
}
with AdS time $t \in [-\infty,\infty]$ in the universal covering space, radius $r>0$ and $\phi \in [0,2\pi]$.

In `hyperbolic or Schwarzschild coordinates' on the other hand, 
\ali{
	\left( \begin{matrix} U \\ V \\ X \\ Y \end{matrix} \right)
	=  \left( \begin{matrix} r \cosh \phi \\ \sqrt{r^2-\lAdS^2} \sinh(t/\lAdS) \\ r \sinh \phi \\  \sqrt{r^2-\lAdS^2} \cosh(t/\lAdS)  \end{matrix} \right)~, \label{SSembcoord}
}
the metric becomes
\ali{
	ds^2 &= -\left(\frac{r^2}{\lAdS^2}-1 \right) dt^2 + \left(\frac{r^2}{\lAdS^2}-1 \right)^{-1} dr^2 + r^2 d\phi^2 \label{BTZsomeversion}
}
or, after the transformation
\equ{
	r \rightarrow r \frac{\lAdS}{r_+}, \quad t \rightarrow t \frac{r_+}{\lAdS}, \quad \phi \rightarrow \phi \frac{r_+}{\lAdS}~,
}
\ali{
	ds^2 &= -\left(\frac{r^2-r_+^2}{\lAdS^2} \right) dt^2 + \left(\frac{r^2-r_+^2}{\lAdS^2} \right)^{-1} dr^2 + r^2 d\phi^2 \qquad (\text{BTZ})~, \label{BTZ}
}
where the range of the hyperbolic angle is  $-\infty < \phi < \infty$.
When referring to this metric as a BTZ black hole, 
it is generally assumed to be the quotient space, with coordinates restricted to the fundamental domain:  $-\pi < \phi < \pi$ . For an unwrapped angle (covering space), $-\infty < \phi < \infty$, we refer to the spacetime as the BTZ black string. 
The metric covers the region outside the horizon $r > r_+$, so this is the 1-sided BTZ black string (it has one conformal boundary at $r \rightarrow \infty$).

Both \eqref{AdSrthetat} and \eqref{BTZ} have the same behaviour near the boundary $r \rightarrow \infty$:
\ali{
ds^2 &\rightarrow \frac{r^2}{\lAdS^2}(-dt^2 + \lAdS^2 d\phi^2) \qquad (r \rightarrow \infty) \\ &= \frac{r^2}{\mathcal R^2}\left(-d\left(\frac{\mathcal R t}{\lAdS}\right)^2 + \mathcal R^2 d\phi^2 \right)~,
}
conformal to
\ali{
	ds^2_{\text{conformal bdy}} &= -d\left(\frac{\mathcal R t}{\lAdS}\right)^2 + \mathcal R^2 d\phi^2~. \label{confbdy}
	}
The conformal factor that was dropped, $r^2/\mathcal R^2$, is completely arbitrary from the boundary CFT point of view, hence the introduction of the arbitrary length scale $\mathcal R$. 

The CFTs can then be said to live at the conformal boundaries of the bulks \eqref{AdSrthetat} and \eqref{BTZ} with rescaled time coordinates $t \rightarrow \mathcal R t/\lAdS$. For AdS, the conformal boundary takes the form of a cylinder $\mathcal C(2\pi \mathcal R)$ with radius $\mathcal R$. After Wick rotation to Euclidean signature, the topology of the BTZ conformal boundary becomes either a cylinder $\mathcal{C(\beta)}$ (for the BTZ string) or a torus $\mathcal T(2\pi\mathcal R, \beta)$ (for the BTZ black hole), with $\beta = 2\pi \lAdS \mathcal R/r_+$ the inverse temperature of the black hole in the bulk.
In the high-temperature limit $\mathcal R/\beta \gg 1/2\pi$  the conformal boundary of the BTZ black hole approaches a cylinder rather than a torus, which corresponds to considering a macroscopic black hole $r_+ \gg \lAdS$. Below the Hawking-Page temperature $\mathcal R/\beta < 1/2\pi$ the dominant saddle-point of the gravity path integral is no longer the BTZ black hole but rather thermal AdS.

\begin{figure}[t!]
	\centering
	\includegraphics[width=4.5in]{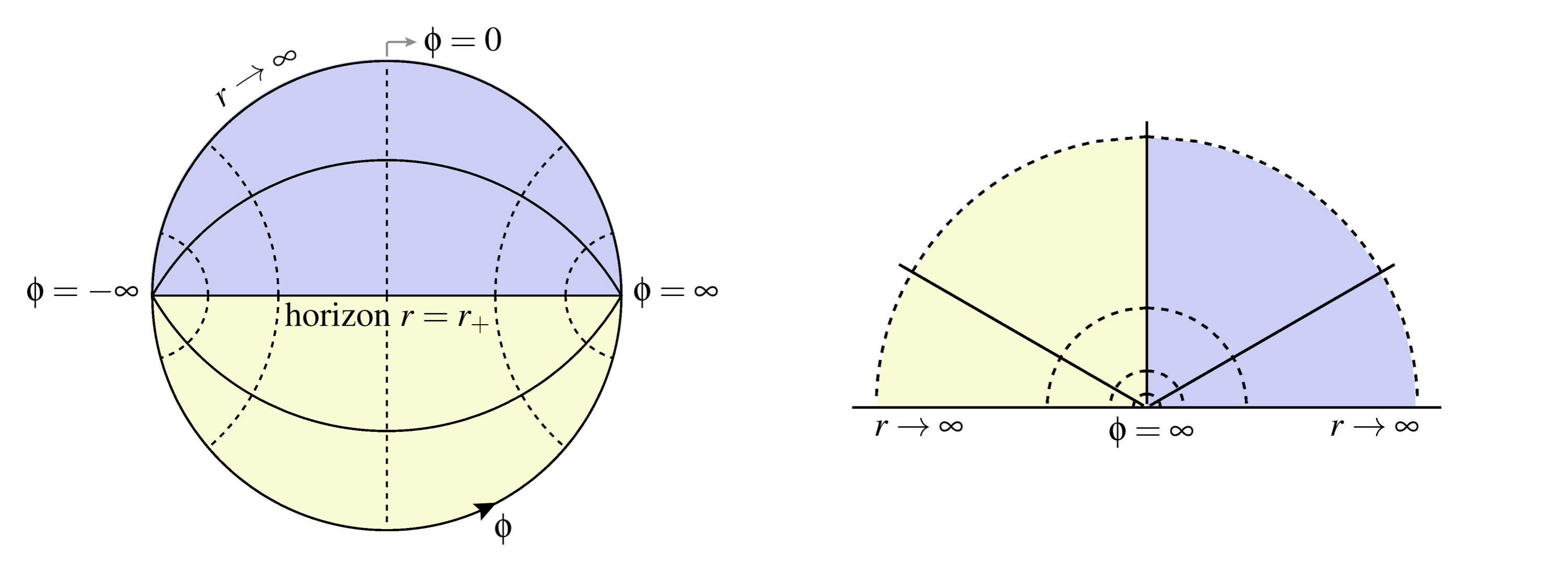}
	\caption{Constant time slice $H_2$ of BTZ, represented as a Poincar\'e disk on the left and the half-plane on the right, showing constant $r$ (solid black) and constant $\phi$ lines (dashed). 
	 The blue region is the region outside the horizon $r>r_+$, covered by the metric \eqref{BTZ} of the 1-sided BTZ. The maximally extended or 2-sided BTZ is obtained when the metric is continued beyond the horizon to include the yellow region. 
		We consider the unquotiented metric $-\infty < \phi < \infty$ in Section \ref{BTZSec}  and the quotiented metric $-\pi < \phi < \pi$ corresponding to a wormhole-like geometry in Section~\ref{BTZqSec}.
}
	\label{PoincareDiskBTZ}
\end{figure}

A constant time slice at $t=0$ of the AdS and BTZ geometries \eqref{AdSrthetat} and \eqref{BTZ} will define an $H_2$ slice $-U^2 +X^2+Y^2=-\lAdS^2$. It can be represented as a Poincar\'e disk, with Poincar\'e coordinates
 \ali{
 	y_1 &= \frac{\lAdS X}{\lAdS+U}(t=0), \quad y_2 = \frac{\lAdS Y}{\lAdS+U}(t=0) \label{Poincarecoord}
 }
that map $r \rightarrow \infty$ to the unit circle. More precisely, the metric of a constant time slice $ds^2_{H_2} = \frac{4 \lAdS^4}{(-\lAdS^2 + y_1^2 + y_2^2)^2} (dy_1^2 + dy_2^2)$
is conformal to the Poincar\'e disk
\ali{
ds^2_{\text{conformal time slice}} = dy_1^2 + dy_2^2~.
}

Another possible representation uses the half-plane coordinates
 \ali{
 	y_1 &= \frac{\lAdS Y}{X+U}(t=0)~, \quad y_2 = \frac{\lAdS^2}{X+U}(t=0)~,
 }
in which the metric becomes $ds^2_{H_2} = \frac{\lAdS^2}{y_2^2} (dy_1^2 + dy_2^2)$.
Both representations are presented in figures~\ref{PoincareDiskAdS} and~\ref{PoincareDiskBTZ}, for AdS and BTZ.

\section{de Sitter Embeddings and Penrose Transformations}\label{EmbeddingPenrose} 

Useful references for the various coordinates on de Sitter space include~\cite{Anninos:2012qw} and~\cite{Mukhanov:991646}.

dS$_2$ is defined as the locus 
\equ{
	-U^2 + X^2 + Y^2 =  \ldS^2 \label{dShyperboloid}
}
in the flat embedding space $R^{1,2}$, with $\ldS$ the dS radius. The induced metric is
\equ{
	ds^2 = -dU^2 + dX^2 + dY^2~.  \label{dSmetric}
}

Global coordinates $(t,\theta)$ with ranges $-\infty \leq t \leq \infty$, $0 \leq \theta \leq 2\pi$ cover the full de Sitter hyperboloid ($-\infty \leq X,Y,U \leq \infty$): 
\be \left( \begin{matrix} X\\Y\\U \end{matrix} \right) = \ldS \left( \begin{matrix} \cosh t \cos \theta\\ \cosh t \sin \theta\\ \sinh t \end{matrix} \right)~. \label{globalembedding}\ee 
The metric \eqref{dSmetric} takes the form \eqref{AdSmetric}. Another time-like global coordinate $\alpha$ is introduced in \eqref{tOfalphaglobal}.

Planar coordinates $(\alpha,\theta)$ with ranges $\alpha > 0, -\infty < \theta < \infty$ cover the \emph{planar patch} $X+U \geq 0$ (illustrated in figure~\ref{PoincareSubregion}b): 
\be \left( \begin{matrix} X\\Y\\U \end{matrix} \right) = \ldS \left( \begin{matrix} \frac{\ldS}{2 \alpha} + \frac{\alpha^2 - \theta^2}{2\ldS \alpha} \\ \frac{\theta}{\alpha} \\ \frac{\ldS}{2\alpha} - \frac{\alpha^2 - \theta^2}{2\ldS \alpha} \end{matrix}\right)~. \label{planarembedding}\ee
The metric \eqref{dSmetric} takes the form \eqref{Kplanar}. 

Hyperbolic coordinates $(\tau,\chi)$ with ranges $0 \leq \tau \leq \infty, -\infty \leq \chi \leq \infty$ cover the region $X > \ldS, -\infty \leq Y \leq \infty, U > 0$, known as the \emph{hyperbolic patch} (figure~\ref{BTZStringSubregion}b): 
\be \left( \begin{matrix} X\\Y\\U \end{matrix} \right) = \ldS \left( \begin{matrix} \cosh \tau \\ \sinh \tau \sinh  \chi \\ \sinh \tau \cosh  \chi  \end{matrix} \right)~. \label{hyperbolicembedding}\ee 
The metric \eqref{dSmetric} takes the form \eqref{BTZeqn}. Other hyperbolic coordinates $\alpha$ and $\theta$ are introduced in \eqref{tOfalphahyp}.  
\\

The corresponding Penrose diagrams (figures \ref{AdSSubregion} and \ref{CSSubregion} in global coordinates, figure \ref{PoincareSubregion} in planar coordinates, and figures \ref{BTZStringSubregion} and  \ref{BTZqSubregion} in hyperbolic coordinates) are obtained through the Penrose transformations 
\ali{
	t_P &= \arctan U~,   \\
	\theta_P &= \arctan\frac{Y}{X}~,
}  
in the respective embedding coordinates. For completeness they are explicitly given below.

In global coordinates 
\ali{
	t_P &= \arctan (\sinh t) \stackrel{\eqref{tOfalphaglobal}}{=} \frac{\pi}{2} - \alpha \qquad (\text{for $0 < \alpha < \pi$})~,   \label{B8} \\
	\theta_P &= \theta \label{B9}~,
}
the metric \eqref{AdSmetric} becomes $ds^2 = (\sec^2 t_P) (-dt_P^2 + d\theta_P^2)$, conformal to $ds^2 = -dt_P^2 + d\theta_P^2$. Note that because of eqs.~\eqref{B8}-\eqref{B9}, the coordinates $t_P$ and $\alpha$, and $\theta_P$ and $\theta$ are used interchangeably when working in global coordinates, namely in the sections on the global AdS and conical geometry kinematic spaces. The global dS$_2$ Penrose diagram is rectangular, with $\theta_P$ the azimuthal angle in a range $-\pi$ to $\pi$ that gets identified. 

\begin{figure}[t!]
	\centering
	\includegraphics[width=6.5in]{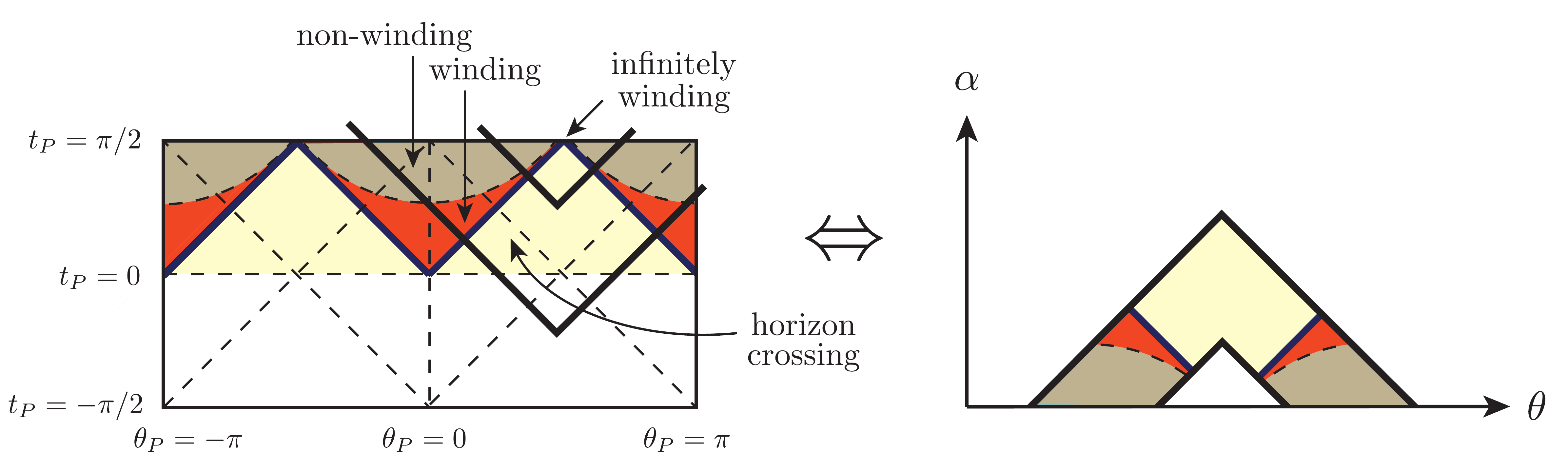}
	\caption{The space of all constant time-slice geodesics of the two-sided BTZ black hole (left), compared with the depiction of it in figure~17 in \cite{Czech:2015kbp} (right). We have shown in thick black lines (on the left) the boundary of the domain depicted in the figure on the right. 
	The coordinates $\alpha$ and $\theta$ on the right are the global de Sitter coordinates of the dS$_2$ whose hyperbolic patch is identified with the BTZ black string kinematic space. They are used only in this instance for comparison, while in the rest of the BTZ discussions in the paper, $\alpha$ and $\theta$ refer to hyperbolic coordinates, see eq.~\eqref{tOfalphahyp}.
}  
	\label{BTZBH2}
\end{figure}

In planar coordinates 
\ali{
	t_P &= \arctan \frac{\ldS^2 + \theta^2 - \alpha^2}{2 \alpha} \label{PlanarPenrose1}~,\\ 
	\theta_P &= \arctan \frac{2 \theta}{\ldS^2 - \theta^2 + \alpha^2} \label{PlanarPenrose2}~.
	}

In hyperbolic coordinates 
\ali{
	t_P &= \arctan(\sinh \tau \cosh \chi) \stackrel{\eqref{tOfalphahyp}}{=} \arccot \left(\sech \frac{2 \pi \theta}{\beta} \sinh \frac{2 \pi \alpha}{\beta} \right)  \label{HyperbolicPenrose1}~,\\
	\theta_P &= \arctan (\tanh \tau \sinh \chi)  \stackrel{\eqref{tOfalphahyp}}{=} \arctan \left(\sech \frac{2 \pi \alpha}{\beta} \sinh \frac{2 \pi \theta}{\beta}\right)\label{HyperbolicPenrose2}~.
}

\section{Two-sided BTZ Black Hole}\label{BTZAppendix}

Maximally extended, asymptotically AdS black holes have two asymptotic regions and these can be identified with two boundary CFTs in an entangled, thermofield state \cite{Israel:1976ur, Maldacena:2001kr}.\footnote{There is some debate about this picture, e.g., \cite{Avery:2013bea, Mathur:2014dia, Chowdhury:2014csa}, but we will assume it is basically correct and explore some implications, without taking a firm stance on the extent of its validity.}
The kinematic space associated with such a two-sided, asymptotically AdS$_3$ black hole 
was discussed in \cite{Czech:2015kbp} and depicted in their figure 17. There, kinematic space is referred to as the space of \emph{all} constant time-slice geodesics, including the winding ones and a set of horizon-crossing geodesics which have one endpoint on each of the two asymptotic regions, while we take the term ``kinematic space'' to refer to the space of Ryu-Takayanagi geodesics only (in a one-to-one mapping between CFT intervals and bulk geodesics). When following their terminology, all the colored regions in our figure \ref{BTZBH1} together have to be used to extend the picture to the two-sided case and compare to the figure in \cite{Czech:2015kbp}. This is done in figure \ref{BTZBH2}.

\acknowledgments
We thank Dionysios Anninos, Alice Bernamonti, Chris Brust, Bartlomiej Czech, Jan de Boer, Netta Engelhardt, Federico Galli, Kurt Hinterbichler, Aitor Lewkowycz, Jennifer Lin, and Herman Verlinde for helpful discussions and input. We especially thank Rachel A. Rosen for discussions and initial collaboration. C.A. is supported in part by the U.S. Department of Energy (DOE) under DOE grant DE-SC0011941, N.C. is supported by the National Science Foundation of Belgium (FWO) Odysseus grant G.0.E52.14N, and C.Z. is supported by NASA ATP grant NNX16AB27G. C.Z. would like to thank the Perimeter Institute Visiting Graduate Fellow program where she learned about kinematic space and tensor networks, and both C.Z. and N.C. would like to thank the New Frontiers in Entanglement Workshop at the University of Pennsylvania for the opportunity to discuss this work and learn about related developments.

\bibliographystyle{JHEP}
\bibliography{references}

\end{document}